\documentclass[reprint,twocolumn,aps,floatfix]{revtex4-1}

\pdfoutput=1
\usepackage[english]{babel}
\usepackage{graphicx}
\usepackage{epsfig}
\usepackage{color}
\usepackage{amssymb}
\usepackage{amsmath}
\usepackage{array}
\usepackage[loose]{units}
\usepackage{hyperref}
\usepackage{cleveref}
\usepackage{braket}
\usepackage[caption=false]{subfig}

\newcommand{\firstpartial}[2]{\frac{\partial #1}{\partial #2}}

\newcommand{\secondpartial}[2]{\frac{\partial^2 #1}{{\partial #2}^2}}

\usepackage{letltxmacro}

\LetLtxMacro{\ORIGselectlanguage}{\selectlanguage}
\makeatletter
\DeclareRobustCommand{\selectlanguage}[1]{%
    \@ifundefined{alias@\string#1}
      {\ORIGselectlanguage{#1}}
      {\begingroup\edef\x{\endgroup
         \noexpand\ORIGselectlanguage{\@nameuse{alias@#1}}}\x}%
}
\newcommand{\definelanguagealias}[2]{%
  \@namedef{alias@#1}{#2}%
}
\makeatother

\definelanguagealias{en}{english}
\newcommand{\UO}{Department of Physics, University of Oregon, Eugene, Oregon}

\begin{document}

\author{Tyler R. Harvey}
\affiliation{\UO}
\author{Benjamin J. McMorran}
\affiliation{\UO}

\title{A Stern-Gerlach-like approach to electron orbital angular momentum measurement}

\date{\today}
                            
\begin{abstract}
  Many methods now exist to prepare free electrons into orbital angular momentum states, and the predicted applications of these electron states as probes of materials and scattering processes are numerous. The development of electron orbital angular momentum measurement techniques has lagged behind. We show that coupling between electron orbital angular momentum and a spatially varying magnetic field produces an angular momentum-dependent focusing effect. We propose a design for an orbital angular momentum measurement device built on this principle. As the method of measurement is non-interferometric, the device works equally well for mixed, superposed and pure final orbital angular momentum states. The energy and orbital angular momentum distributions of inelastically scattered electrons may be simultaneously measurable with this technique. 
\end{abstract}

\maketitle
How does one measure the orbital angular momentum (OAM) of the quantum state of an unbound, massive, charged particle after interaction with another particle or a material? Free electrons with OAM, also called electron vortices, are now routinely prepared in electron microscopes \cite{uchida_generation_2010,verbeeck_production_2010,mcmorran_electron_2011,saitoh_production_2012,schattschneider_novel_2012,blackburn_vortex_2014,beche_magnetic_2014,harvey_efficient_2014,grillo_highly_2014,shiloh_sculpturing_2014,beche_efficient_????}, and control of this new degree of freedom is widely recognized as a useful tool in the both the study of materials and basic physical processes \cite{verbeeck_production_2010,schattschneider_mapping_2012,asenjo-garcia_dichroism_2014,harvey_demonstration_2015}. A variety of impressive techniques now exist to prepare an electron in an OAM state. Full control of free electron orbital angular momentum, though, demands good measurement tools.

One of the most promising potential applications of electron OAM--measurement of magnetization at atomic resolution via helical dichroism spectroscopy--serves as an excellent example of the importance of both preparation and post-selection in applications of electron OAM. Magnetic dichroism has, surprisingly, not yet been realized with electrons prepared in orbital angular momentum states. This application is analogous to X-ray Magnetic Circular Dichroism (XMCD), a widely-used technique for magnetization measurement based on the ratios of core-transition peaks in left- and right-circularly polarized X-ray absorption spectra. There exists a crucial difference, though, between circular dichroism--which involves controlled transfer of photon spin angular momentum--and helical dichroism--which involves controlled transfer of electron orbital angular momentum \cite{harvey_demonstration_2015}. Photons are massless and can be absorbed by materials, so the final state of a photon in a circular dichroism measurement is just the vacuum state. Electrons are massive, and carry away non-zero energy and angular momentum from an interaction. If we seek to gain the most information about a material in an electron spectroscopy experiment, we ought to measure both the final electron energy and OAM \footnote{Simulations \cite{rusz_achieving_2014} suggest that a small dichroism effect does exist when one measures only the probability density of the final state--and therefore traces out OAM in the final state.}. Helical dichroism can be made far more efficient with careful post-selection of electron OAM states. This insight, in fact, applies to many applications of electron OAM.

There are a wide range of applications of good OAM post-selection. Theoretical predictions and simulation suggest that electron impact ionization \cite{macek_theory_2010,ward_effect_2014}, photoionization \cite{ngoko_djiokap_electron_2015,geneaux_attosecond_2015}, electron-atom scattering \cite{serbo_scattering_2015}, material investigation with angle-resolved photoelectron spectroscopy \cite{takahashi_berry_2015} and electron energy loss spectroscopy \cite{schuler_disentangling_2015}, production of spin-polarized electrons \cite{karimi_spin--orbital_2012}, and even high-energy elementary particle collisions \cite{ivanov_measuring_2012} can produce non-trivial final OAM states and could therefore benefit from OAM post-selection.

Several techniques have so far been developed for electron OAM measurement; they work well as quality-assurance tests for new orbital angular momentum state preparation techniques. All have limitations that prohibit their application to post-selection of a single final state of an inelastic interaction. 
Self-interferometric techniques \cite{guzzinati_measuring_2014,clark_quantitative_2014,shiloh_unveiling_2015} depend on analysis of the spatial distribution an electron after a transformation. In general, inelastic interaction of an electron and a material produces mixed electron final states thanks to entanglement with the material. Mixed and superposed OAM states are extremely difficult to quantitatively measure with self-interferometric techniques \cite{shutova_measurement_2016}. Furthermore, energy-filtered TEM is necessary to isolate and analyze the spatial distribution of the states scattered to a given energy. Holographic phase-flattening \cite{yahn_addition_2013,saitoh_measuring_2013} can partially spatially isolate a single component of a mix of inelastically scattered final OAM states, but is currently terribly inefficient.

We propose a technique for OAM post-selection based on coupling of OAM to a spatially varing magnetic field. The effect is analogous to the coupling between spin and a spatially varying magnetic field that Stern and Gerlach employed in their demonstration of the quantization of spin \cite{gerlach_experimentelle_1922}. In the Stern-Gerlach device, spins aligned (anti-aligned) with the magnetic field are pulled by the Zeeman interaction toward the side of the device with higher (lower) field strength. Unlike the Stern-Gerlach device for measurement of spin, we consider a cylindrically symmetric design for measurement of OAM. Cylindrical symmetry gaurantees conserve electron OAM through the measurement device \cite{noether_invariant_1971} and control the Lorentz force \cite{batelaan_stern-gerlach_1997}. Fortunately, cylindrically symmetric, spatially varying magnetic fields find great use as electron round lenses \cite{reimer_transmission_2008}. We show that the coupling of OAM to the field of a magnetic round lens produces a shift in the focal length of a magnetic round lens. In this proposed device, electrons with orbital angular momentum aligned (anti-aligned) with the magnetic field are pushed outward away from (pulled inward toward) the strong magnetic field along the optic axis, as the electron charge is negative.




For a state propagating along the $z$-axis, the transfer function of a lens with focal length $f$ on an electron with wavelength $\lambda$ is
\begin{equation} \label{eq:lens_transfer}
  U_{\textrm{lens}} = e^{-i\frac{\pi\rho^2}{\lambda f}}
\end{equation}
where $\rho$ is distance from the $z$-axis in cylindrical coordinates $(z,\rho,\phi)$.

If instead we want an OAM-dependent focal length, we'll want to construct a transfer function
\begin{equation} \label{eq:OAM_lens_transfer}
  U = \exp\left(-i \frac{L_z \rho^2}{\hbar b^2}\right).
\end{equation}
This transfer function produces a quantum non-demolition measurement of orbital anguar momentum: OAM is an eigenstate of both this transfer function and the free-space Hamiltonian. The effect of this transfer function on an orbital angular momentum state is visualized in Figure \ref{fig:phase_effect} in the Supplemental Material.

\begin{figure}[thb]
  \subfloat[]{\label{subfig:superposition:1}
  \includegraphics[width=0.3\columnwidth]{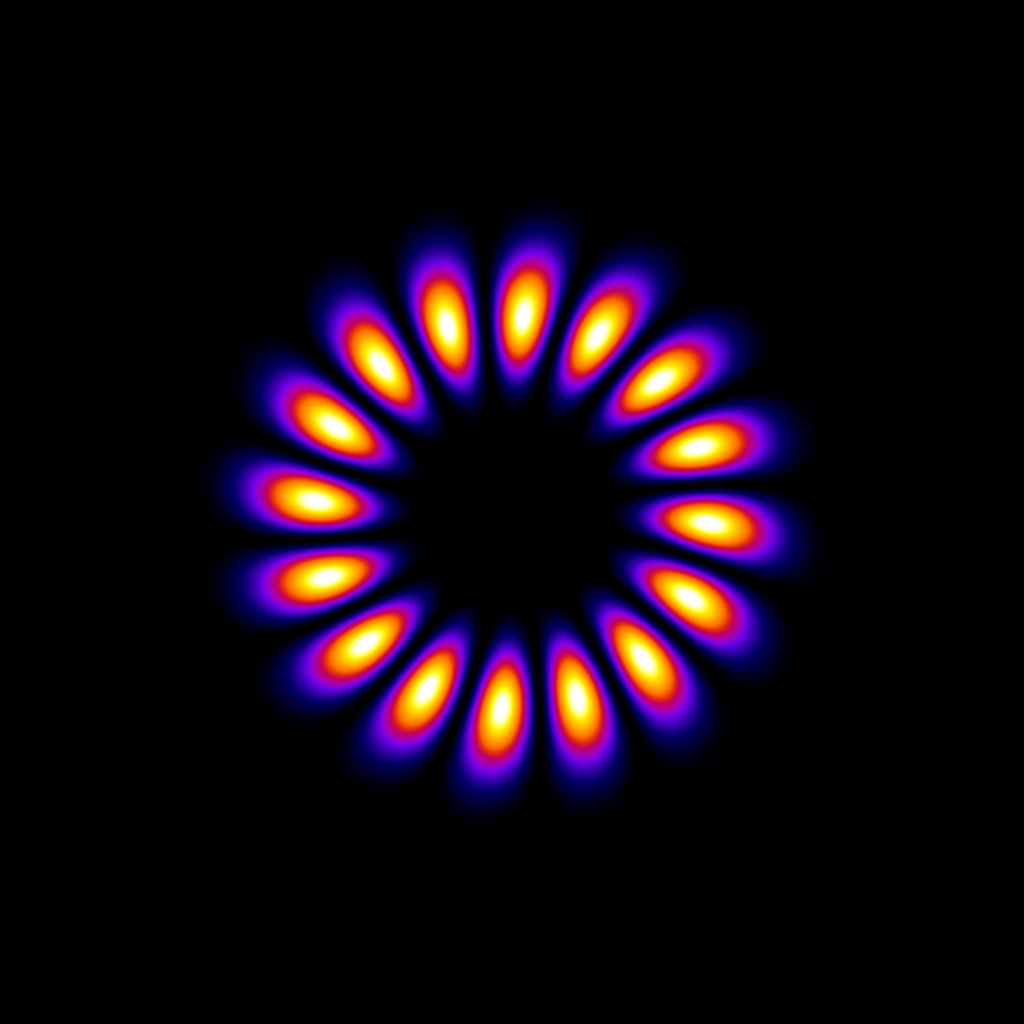} }
  \subfloat[]{\label{subfig:superposition:2}
  \includegraphics[width=0.3\columnwidth]{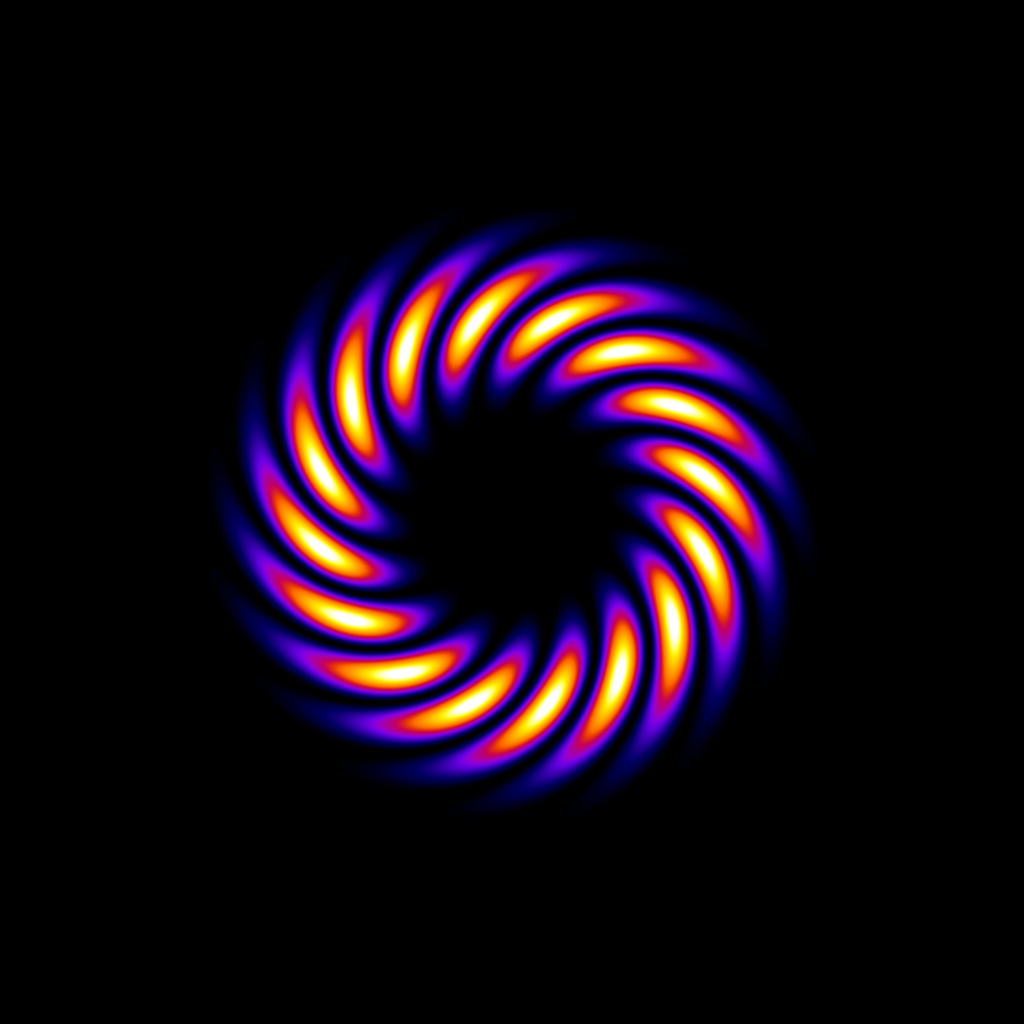} }
  \subfloat[]{\label{subfig:superposition:3}
  \includegraphics[width=0.3\columnwidth]{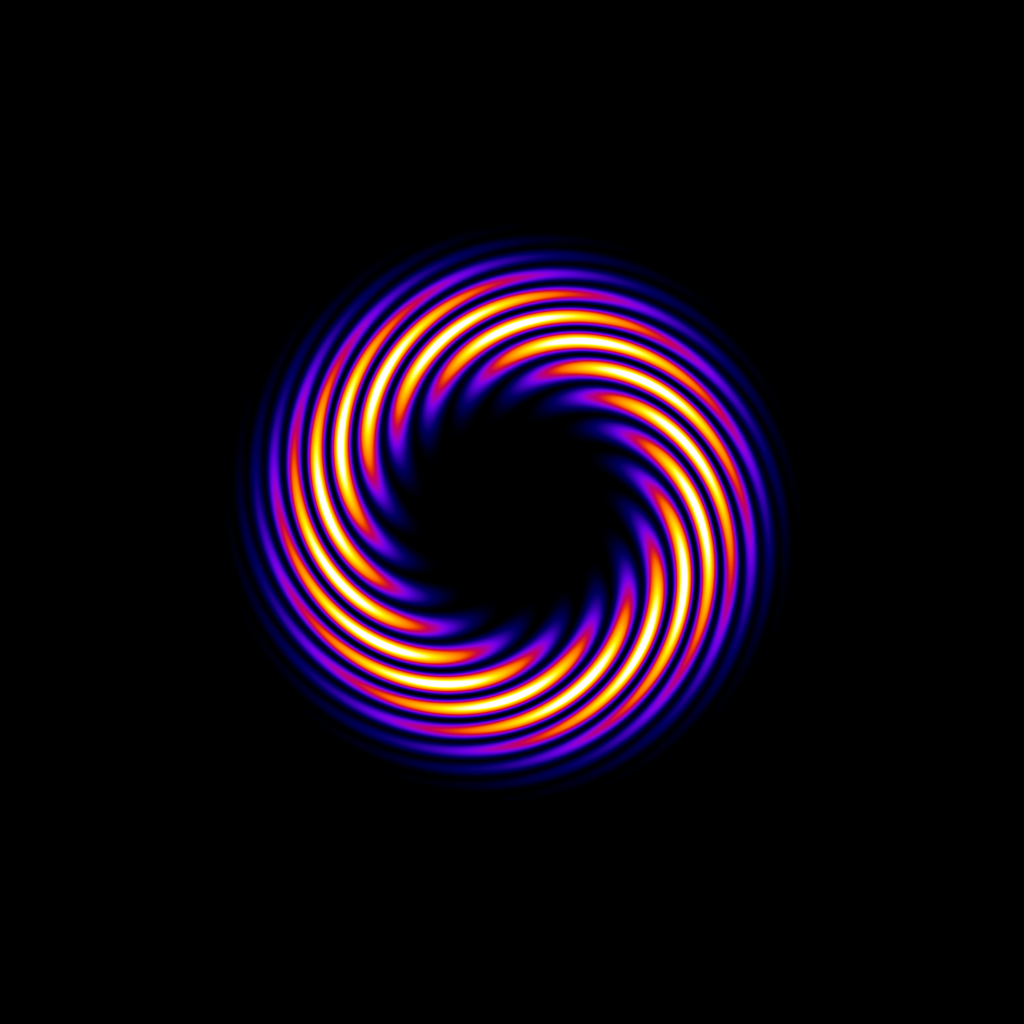} } \\
  \subfloat[]{\label{subfig:superposition:4}
  \includegraphics[width=0.3\columnwidth]{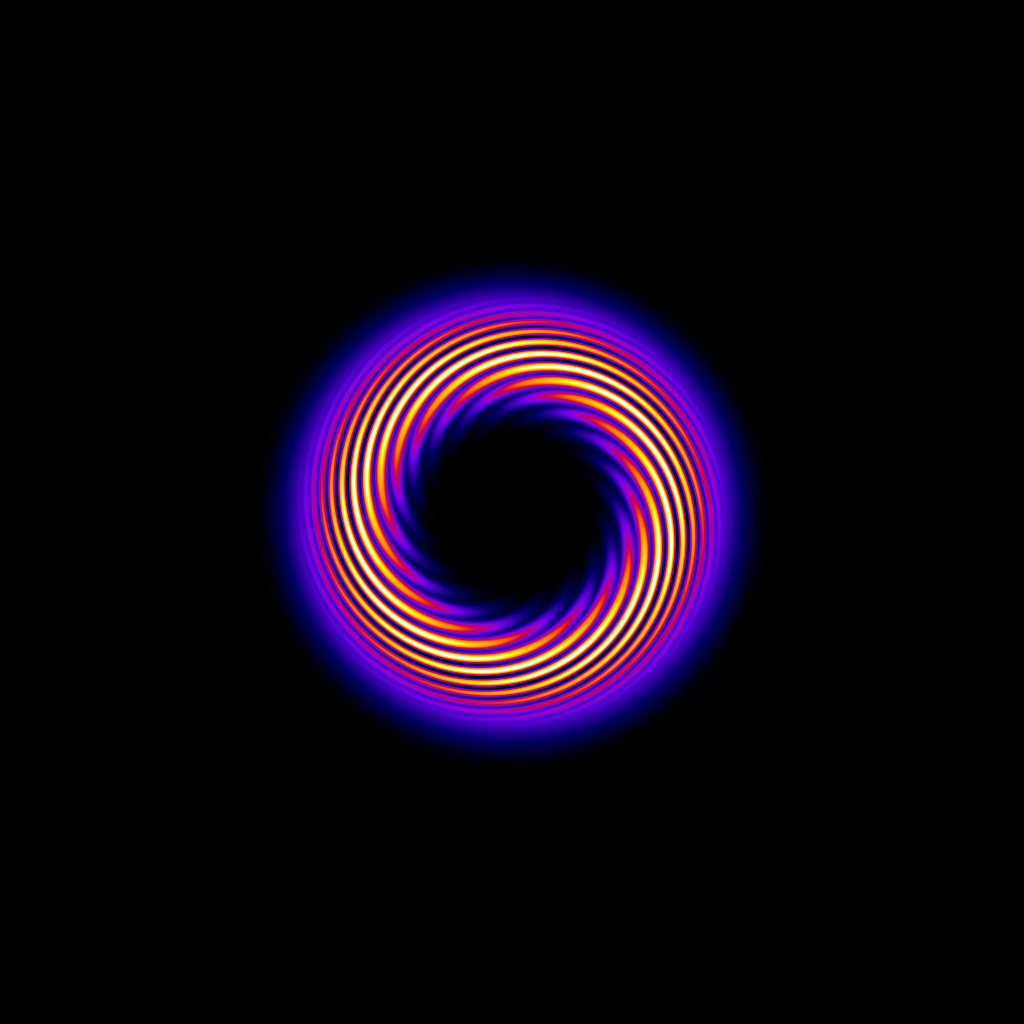} }
  \subfloat[]{\label{subfig:superposition:5}
  \includegraphics[width=0.3\columnwidth]{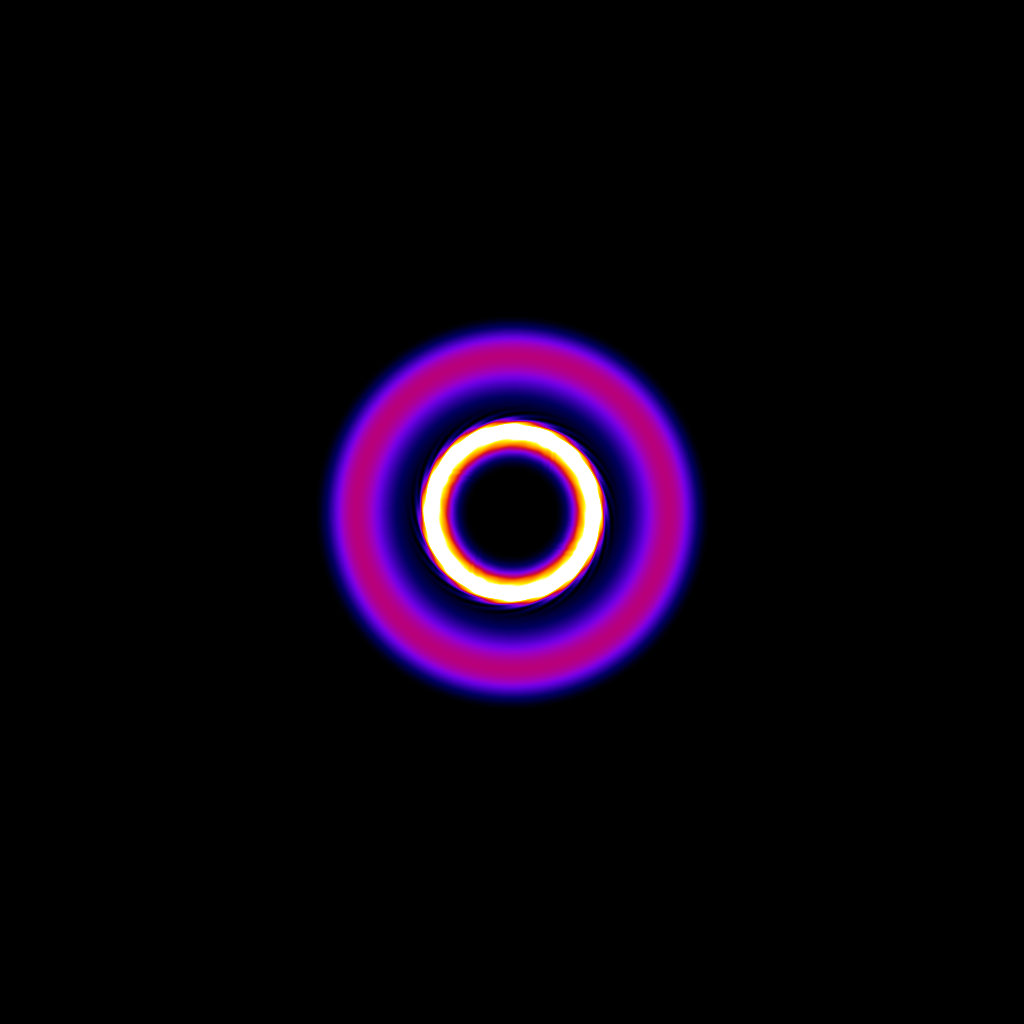} }
  \subfloat[]{\label{subfig:superposition:6}
  \includegraphics[width=0.3\columnwidth]{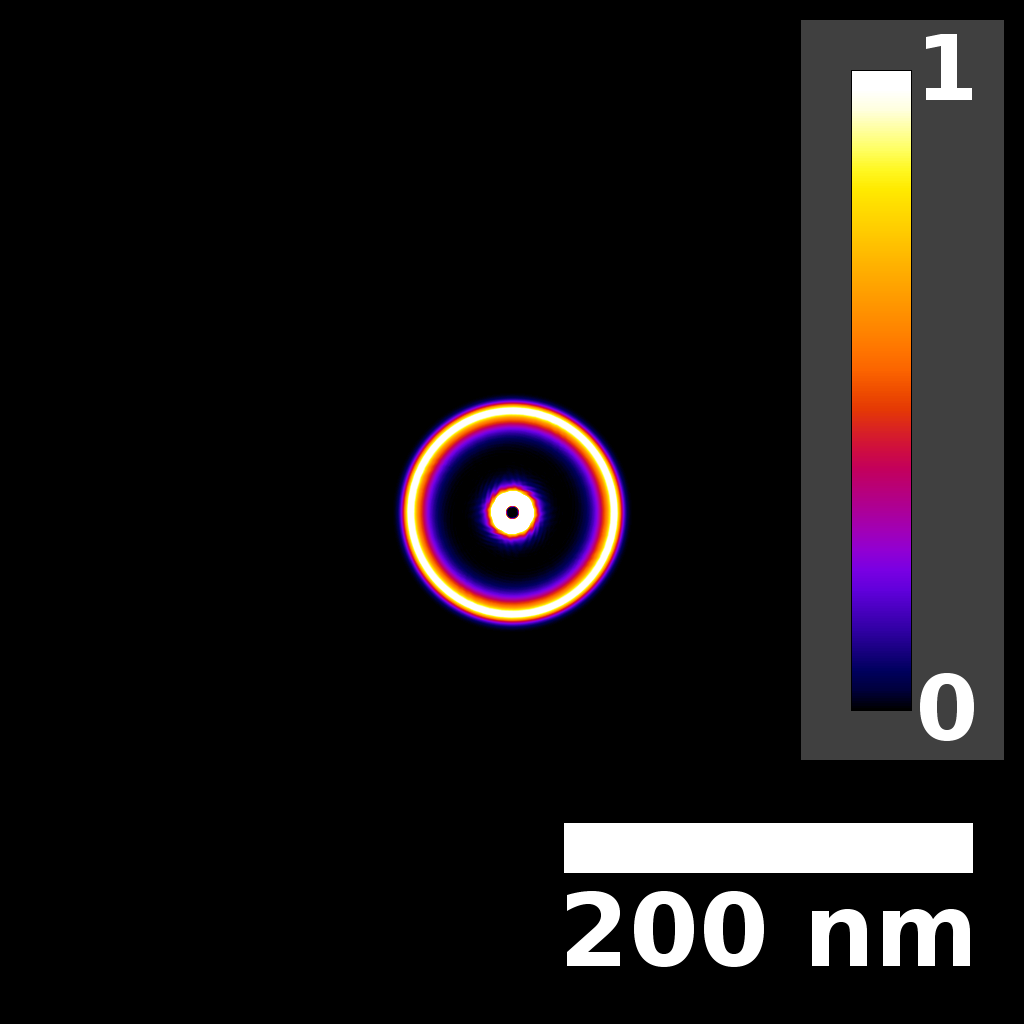} } \\
  \subfloat[]{\label{subfig:basictrajectory}
  \includegraphics[width=\columnwidth]{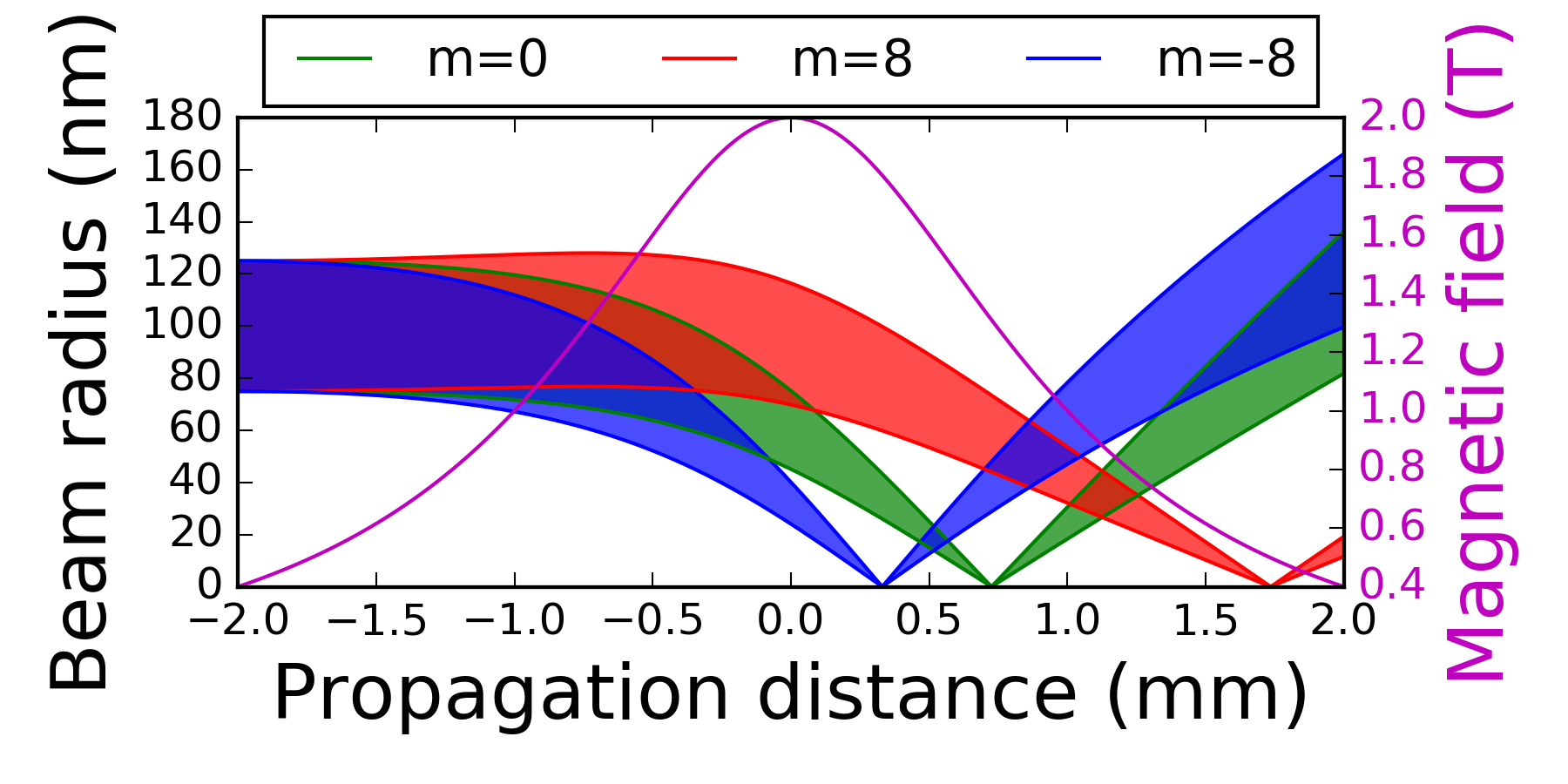} }
  \caption{(top) Multislice-simulated intensities of a superposition of $m=\pm8$ Laguerre-Gaussian orbital angular momentum modes in a Glaser-model, i.e. equation \eqref{eq:glaser}, field with maximum field strength $B_0 = \unit[2]{\textrm{T}}$, longitudinal extent $a=\unit[1]{\textrm{mm}}$ and an OAM dispersion length $b = \unit[79]{\textrm{nm}}$, sampled at (a) $\unit[-2.0]{\textrm{mm}}$, (b) $\unit[-1.5]{\textrm{mm}}$, (c) $\unit[-1.0]{\textrm{mm}}$, (d) $\unit[-0.5]{\textrm{mm}}$, (e)  $\unit[0.0]{\textrm{mm}}$ and (f) $\unit[0.5]{\textrm{mm}}$ from the center of the lens. (bottom) Ray trajectories for $m=-8$, $m=0$ and $m=+8$ modes calculated by numerical integration of the radial equation of motion corresponding to the full Hamiltonian in equation \eqref{eq:fullH} with a Glaser-model field. \label{fig:basicsimulation}}
\end{figure}

With a careful study of the link between terms in the electron Hamiltonian and the resulting transfer function (see section \ref{sect:H_lens} of the Supplemental Material), we can see that we'll produce a transfer function like \eqref{eq:OAM_lens_transfer} with the vector potential
\begin{equation} \label{eq:modelvp}
  \mathbf{A} = \left(B_1(z)\frac{\rho}{2}-B_3(z)\frac{\rho^3}{8 b^2}\right)\hat{\boldsymbol{\phi}}.
\end{equation}
where $B_1$ and  $B_3$ describe the longitudinal profile of the field. This vector potential corresponds to a magnetic field that points along the $\pm\hat{\mathbf{z}}$-direction at the origin and curves outward away from the origin over a length scale $b$. We'll call $b$ the dispersion length. The corresponding Hamiltonian for an electron in this vector potential includes two lensing terms,
\begin{equation}
  H_{\textrm{lens}} = \frac{1}{8 m_e}\left(e^2 B_1^2 - \frac{e B_3 L_z}{b^2}\right)\rho^2.
\end{equation}
where $e=|e|$ is the magntide of the electron charge. The latter term produces an orbital angular momentum-dependence in the focal length of the lensing effect. Figure \ref{fig:basicsimulation} illustrates this lensing effect with multislice-simulated \cite{grillo_quantum_2013} propagation of superposed orbital angular momentum states in this Hamiltonian \footnote{The non-unitarity of the lowest-order approximation to the transformation induced by $L_z$ terms produces a small, unphysical loss of probability density with each slice.} and ray trajectories calculated by numerical integration of the radial equation of motion.

The vector potential \eqref{eq:modelvp} above is an approximation to the vector potential of any cylindrically symmetric current distribution with azimuthal current flow. The dispersion length is related to the radial extent of the current distribution.  In the case of a single loop of wire of radius $R$, $b$ is just $R$. So, in fact, there exists a small OAM-dependence in the focal length of any standard magnetic round lens. The key to designing an orbital angular momentum measurement device is to isolate or maximize the OAM-dependence.

To do anything with an orbital angular momentum-dependent lensing effect, we need to know the focal length of the lens. As we show in detail in section \ref{sect:paraxSch} of the Supplemental Material, an initially collimated eigenstate of $L_z$ with quantum number $m$ that passes through the vector potential \eqref{eq:modelvp} will be focused at a distance from the center of the potential
\begin{equation} \label{eq:thin_f}
  \frac{1}{f_m} = \frac{e^2}{8 m_e E} \int_{-\infty}^{\infty} B_1^2(z) - \frac{m \hbar}{e b^2} B_3(z) \textrm{d}z.
\end{equation}
We can more simply rewrite this as
\begin{equation}
  f_m = \frac{f_0}{1-\Lambda m},
\end{equation}
where $f_0$ is the focal length of the $m=0$ eigenstate and the OAM dispersion coefficient $\Lambda = \frac{\beta_0 \hbar}{eB_0 b^2}$ is a dimensionless constant that depends only on the peak field strength $B_0$, the dispersion length, $b$, fundamental constants, and an $O(1)$ number $\beta_0$ that depends on the shape of the current distribution. We calculate this focal length for several current distributions in section \ref{sect:physical_current} of the Supplemental Material. For small OAM dispersion $\Lambda$, therefore, focal length is approximately linear with OAM.
\begin{equation} \label{eq:linear_f}
  f_m \approx f_0\left(1+\Lambda m\right).
\end{equation}
\begin{figure}[ht]
  \includegraphics[width=0.5\columnwidth]{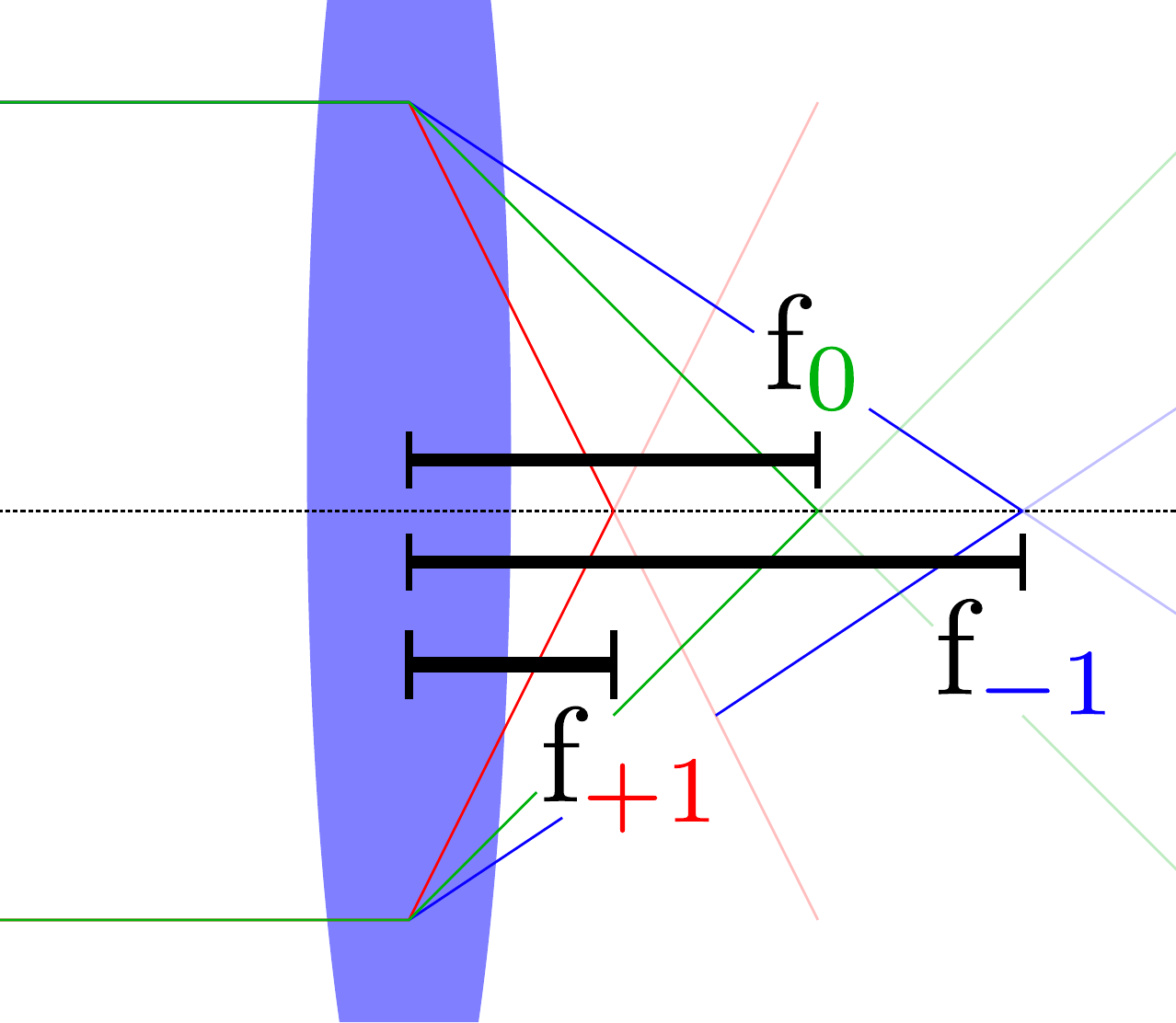}
  \caption{Ray diagrams for a lens (blue disk) with a strongly OAM-dependent focal length $f_m$ as given in \eqref{eq:linear_f}. (red) rays for $m=+1$ electrons; (green) rays for $m=0$ electrons; (blue) rays for $m=-1$ electrons. \label{fig:f_m}}.
\end{figure}
When the current source for the vector potential in \eqref{eq:modelvp} is a superconducting ring, there's an easy physical interpretation of the OAM dispersion coefficient $\Lambda$. A superconducting ring of radius $b$ encloses an area $\pi b^2$ and has an OAM dispersion coeffifient inversely proportional to the number $n$ of flux quanta in the ring, as $n \propto \frac{B_0 \pi b^2}{\Phi_0}$ and the flux quantum $\Phi_0 = \frac{h}{2e}$. We can therefore write the focal length of a lens made of superconducting ring as
\begin{equation}
  f_m \approx f_0 \left(1+\beta_1 \frac{m}{n}\right)
\end{equation}
where $\beta_1$ is another $O(1)$ number.

\begin{figure}[ht] 
  \includegraphics[width=\columnwidth]{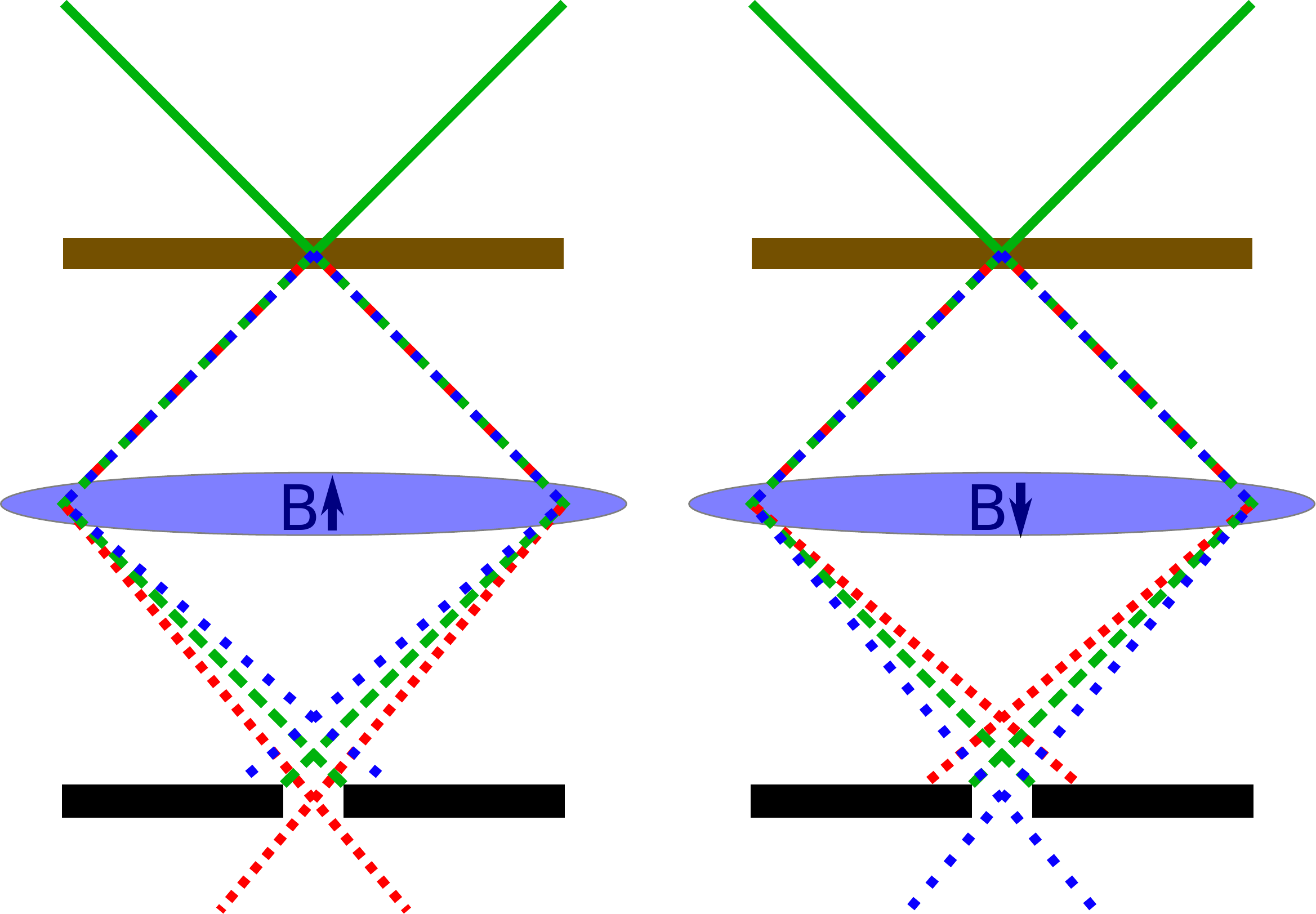}
  \caption{Schematic ray diagrams for a dichroism experiment based on the OAM-dependent lensing effect. Interaction with a specimen (brown) produces a mix of outgoing $m=+1$ (red), $m=0$ (green) and $m=-1$ (blue) orbital angular momentum eigenstates. 
    (left) An aperture (black) preferentially admits the $m=+1$ OAM eigenstate. The $m=+1$ state has a longer focal length in the positive-polarity lens (blue disk). 
  (right) The aperture preferentially admits the $m=-1$ OAM state when the lens polarity is flipped. \label{fig:lens_flip_dichroism}}
\end{figure}

If a measurement device can be constructed with a large OAM dispersion coefficient $\Lambda \sim 1$, the simplest application of this lensing effect needs only a small aperture to select out one focused mode in the appropriate plane, as shown in Figure \ref{fig:lens_flip_dichroism}. This lensing effect makes possible a straightforward helical dichroism experiment without any need for incident OAM. One can see that the focal length \eqref{eq:thin_f} has an OAM-independent part that depends on the magnitude of the lensing magnetic field and an OAM-dependent part that depends on the sign and direction of the magnetic field. In other words, one can control OAM dispersion via the direction of the lensing field. With an aperture set to preferentially admit the $m=+1$ mode, one can flip the polarity of the lens and therefore flip the sign of OAM dispersion and instead admit the $m=-1$ mode without physically moving anything. This experiment likely will require an exceptionally stable microscope and careful alignment to ensure that no other beam properties change upon a lens polarity flip. 

Several physical sources could produce a magnetic field with a significant OAM dispersion. The most obvious, but perhaps the most difficult to build, is a nanoscale solenoid. A solenoid with a radius on the order of \unit[100]{nm} and a peak magnetic field on the order of \unit[1]{Tesla} produces an OAM dispersion coefficient on the order of 0.1. The bound current density on the surface of a hole in an out-of-plane-polarized ferromagnetic thin film looks identical to the current density of a solenoid and could produce the same dispersion; such a hole would be far more easily nanofabricated and has the advantage over a loop of wire that the normal lensing effect will be partially canceled in the hole. A pulsed laser with a radially polarized magnetic field has the appropriate symmetry. The laser used in a recent experiment to prepare well-defined electron momentum states \cite{feist_quantum_2015}, with a peak magnetic field of \unit[0.334]{T} and a spot size of ~\unit[50]{um}, would produce an OAM dispersion coefficient on the order of $10^{-6}$. This might be improved by several orders of magnitude with plasmonic field enhancement.

A completely orthogonal approach to realization of an OAM measurement device of this kind might involve stacking many lenses with a small OAM dispersion coefficient in a manner that magnifies the OAM-dependent effect. We discuss two possible designs for a stacked lens OAM measurement device in sections \ref{sect:stacked1} and \ref{sect:stacked2} of the Supplemental Material. Both designs produce strongly OAM-dependent magnification but only weakly OAM-dependent image plane locations.

\begin{figure}[ht]
  \includegraphics[width=\columnwidth]{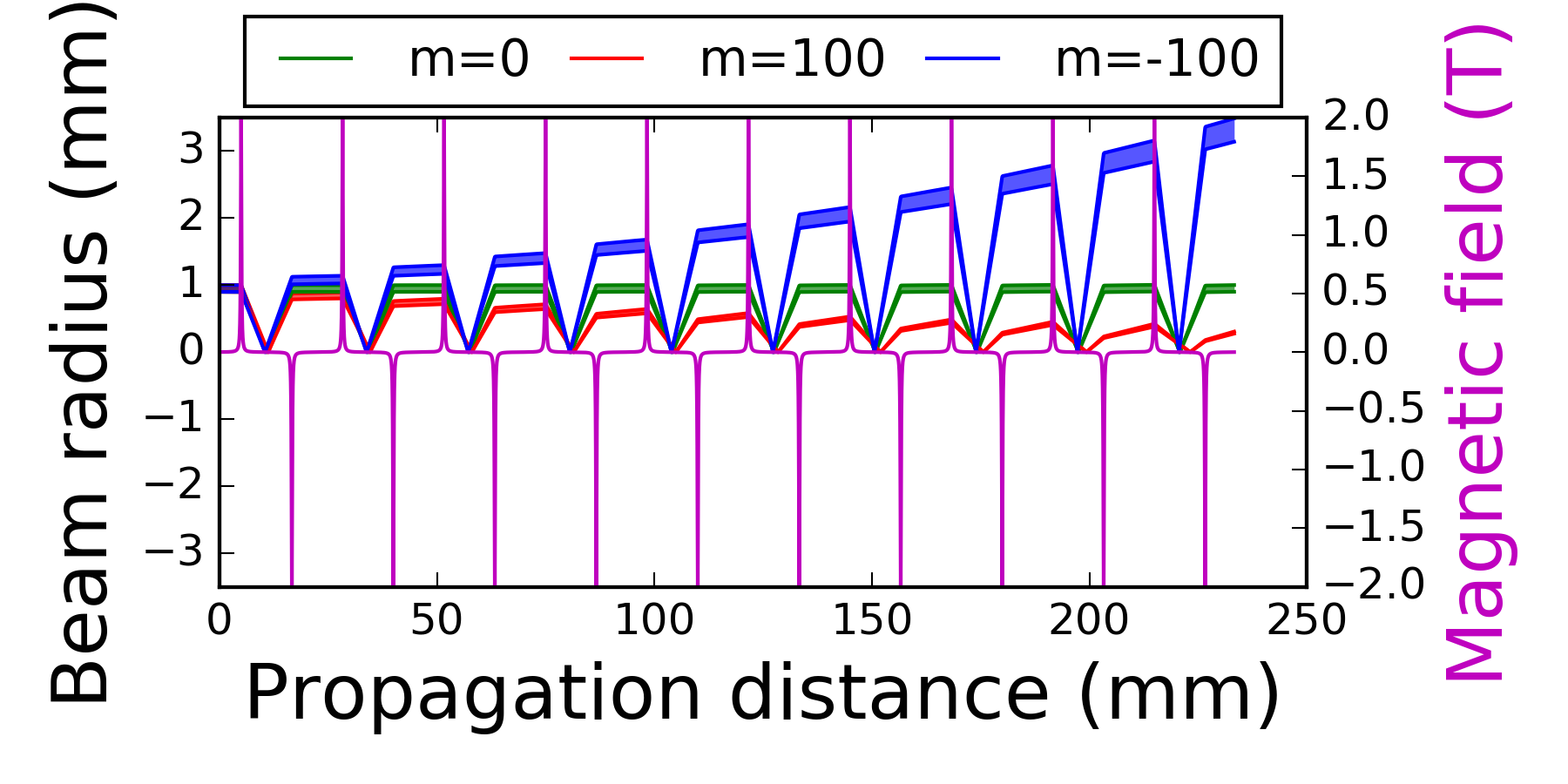}
  \caption{Ray trajectories of $m=-100$ (blue), $m=0$ (green) and $m=+100$ (red) orbital angular momentum modes propagating in a set of ten stacked afocal systems \eqref{eq:mag_afocal} of Glaser-model lenses \eqref{eq:glaser} with longitudinal extent $a = \unit[100]{\mu \textrm{m}}$, OAM dispersion length $b = \unit[1]{\mu \textrm{m}}$ and maximum field strength $B_0 = \unit[2]{\textrm{T}}$. The magnification of OAM goes exponentially with the number of lens sets. \label{fig:afocal} }
\end{figure}

We have demonstrated a Stern-Gerlach-like effect for measurement of electron orbital angular momentum. The measurement technique is applicable to the mixed states produced by inelastic scattering, which are otherwise difficult to measure. We outlined several strategies for practical implementation of this measurement device. If the device can be successfully built and integrated into electron spectrometers, simultaneous measurement of electron energy and orbital angular momentum distributions may be possible.

\begin{acknowledgements}
  We thank Vincenzo Grillo for many helpful conversations, ideas, support for multislice simulations and a careful reading of the manuscript. We appreciate careful scrutiny of the manuscript from Fehmi Yasin and Jordan Pierce. Lastly, T.R.H. thanks Jordan Chess for his ever-present willingness to critically engage with a new idea.
\end{acknowledgements}

\newpage
\begin{widetext}
  \section{Supplemental Material}

\section{Hamiltonian approach to standard magnetic lensing \label{sect:H_lens}}

How does standard magnetic lensing work? Let's schematically identify terms in the electron Hamiltonian that cause lensing by inspecting the time evolution operator that results from the Hamiltonian. Although this approach won't get us the focal length of a magnetic lens--we'll need to solve the paraxial Schrodinger equation (see section \ref{sect:paraxSch}) to do that--but it will allow us some intiutive insight with regard to lensing behavior. 

In order to construct a time evolution operator that causes lensing,
\begin{equation}
  U(t) = \exp\left(iHt/\hbar\right) \propto \exp\left(-i \frac{\pi \rho^2}{\lambda f }\right)
\end{equation}
we need a Hamiltionian with a $\rho^2$ term. The round magnetostatic lenses used most frequenty used in electron microscopes have a vector potential
\begin{equation} \label{eq:ideal_round_lens_A}
  \mathbf{A} = \frac{B_0(z)}{2}\rho \hat{\boldsymbol{\phi}}
\end{equation}
In cartesian coordinates, we see
\begin{align}
  \mathbf{A} &= \frac{B_0(z)}{2}\left(x\hat{\mathbf{y}}-y\hat{\mathbf{x}}\right) \\
  \mathbf{A}\cdot\mathbf{p} &= \frac{B_0}{2}\left(xp_y-yp_x\right) = \frac{B_0}{2} L_z \\
  A^2 & = \frac{B_0^2(z) \rho^2}{4}
\end{align}
which produces a non-relativistic Hamiltonian
\begin{align}
  H &= \frac{1}{2m_e}\left(\mathbf{p}+e\mathbf{A}\right)^2= H_0 + H_{\textrm{Larmor}}+H_2\\
  H_0 &= \frac{p^2}{2 m_e} \\
  H_{\textrm{Larmor}} &= \frac{e B_0 L_z}{2 m_e} \\
  H_2 &= \frac{e^2 B_0^2(z)}{8 m_e}\rho^2
\end{align}
where the electron charge $q_e = -e$, $H_0$ is the free Hamiltonian, $H_{\textrm{Larmor}}$ causes image rotation through a magnetic lens, and $H_2$ causes lensing.

\section{Full Hamiltonian}

Let now add in a lowest-order radial correction to the vector potential that represents the finite size of the current source. In the main text, we wrote our model vector potential as eq. \eqref{eq:modelvp},
\begin{equation} 
  \mathbf{A} = \left(B_1(z)\frac{\rho}{2}-B_3(z)\frac{\rho^3}{8 b^2}\right)\hat{\boldsymbol{\phi}}.
\end{equation}
The full non-relativistic Hamiltonian for this vector potential is 
\begin{align} \label{eq:allH}
  H &= H_0 + H_{\textrm{Larmor}} + H_2 + H_4 + H_6 \\
  H_0 &= \frac{p^2}{2 m_e} \\
  H_{\textrm{Larmor}} &= \frac{e B_1 L_z}{2 m_e} \\
  H_2 &= \frac{1}{8 m_e}\left(e^2 B_1^2 - \frac{e B_3 L_z}{b^2}\right)\rho^2 \\ \label{eq:H2}
  H_4+H_6 &= \frac{e^2 }{8 m_e}\left(-B_1B_3\frac{\rho^4}{2 b^2}+B_3^2\frac{\rho^6}{16 b^4}\right)
\end{align}
where, as in the lowest-order description, $H_0$ is the free Hamiltonian, $H_{\textrm{Larmor}}$ causes image rotation. $H_4$ and $H_6$ are higher-order terms that traditional multipole magnetic corrector elements can cancel without affecting $H_2$ (see Sect. \ref{sect:multipolar_corrector_A}). The term we care about is $H_2$, which again produces lensing and now has two contributions: the standard magnetic lensing term, $\frac{e^2 B_1^2}{8 m_e}\rho^2$ \cite{reimer_transmission_2008}, and the OAM-dependent term $\frac{eB_3 L_z}{8 m_e b^2}\rho^2$. Just as the Larmor term can be interpreted either as causing a rotation or an OAM-dependent phase shift, the OAM-depent lensing term could equivalently be interpreted as a radius-dependent rotation. This term is one source of spiral distortion, an aberration that Scherzer documented in 1937 \cite{marai_scherzers_1976}.

If higher-order terms $H_4+H_6$ are corrected (see section \ref{sect:multipolar_corrector_A}), the full orbital angular momentum measurement Hamiltonian is
\begin{equation} \label{eq:fullH}
  H_0 = \frac{p^2}{2 m_e} + \frac{e B_1 L_z}{2 m_e} + \frac{1}{8 m_e}\left(e^2 B_1^2 - \frac{e B_3 L_z}{b^2}\right)\rho^2 
\end{equation}

\begin{figure}[h]
  \includegraphics[width=\columnwidth]{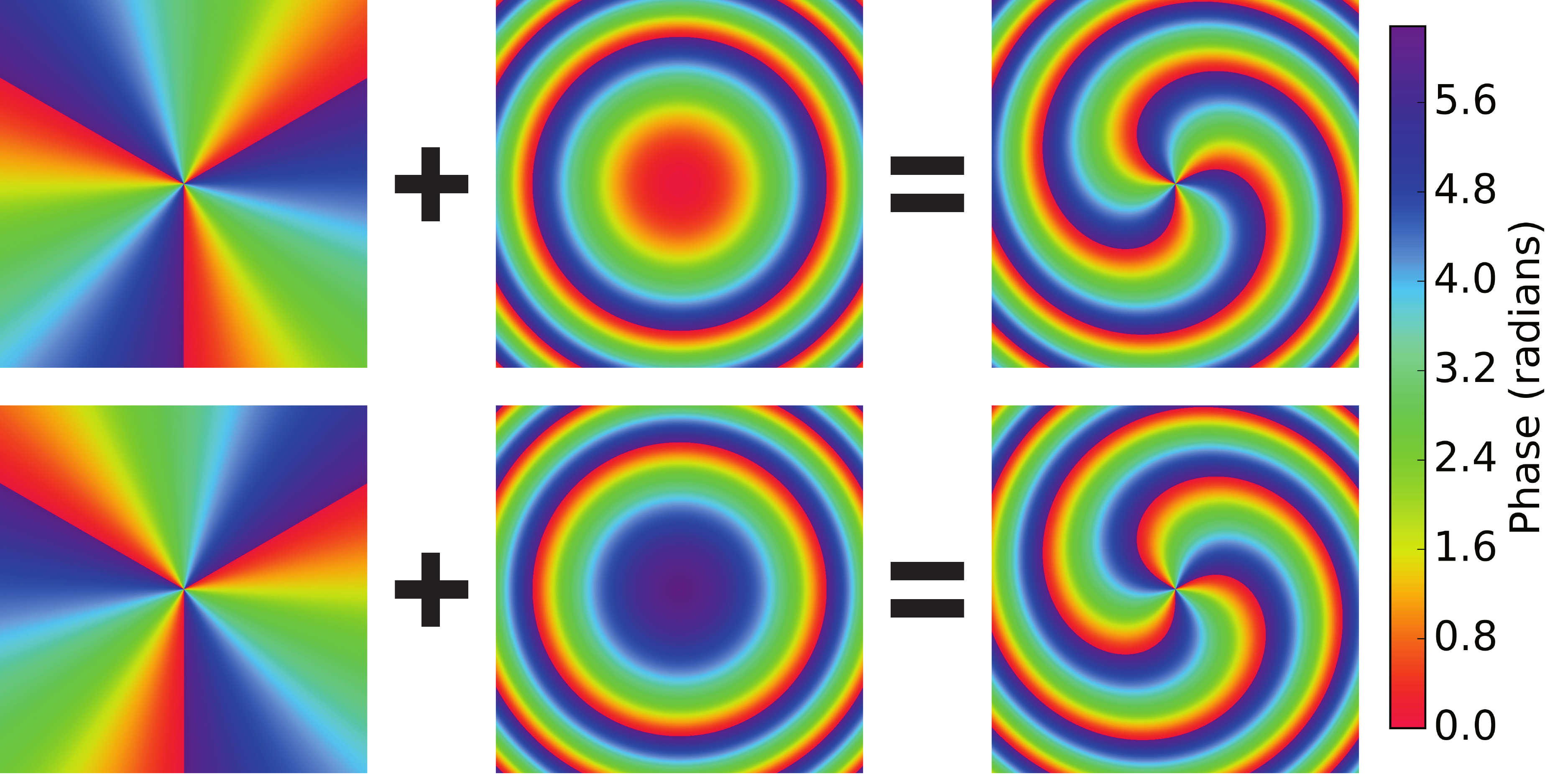}
  \caption{Illstration of the effect of the orbital angular momentum-dependent lensing effect. Propagation of a wave with orbital angular momentum--and therefore an azimuthal phase (first column)--in a Hamiltonian with the orbital angular momentum-dependent lensing term in equation \eqref{eq:H2} produces a parabolic phase (second column) in proportion to the orbital angular momentum. The result is a spiralling phase with a winding magnitude and direction that depends on orbital angular momentum (third column). \label{fig:phase_effect} }
\end{figure}

\section{Calculation of thin lens focal length through solution of paraxial Schrodinger equation \label{sect:paraxSch}}

In this section, we derive an expression for the orbital angular momentum-dependent focal length \eqref{eq:thin_f} for a thin magnetostatic lens. One can use the same formalism for the focal length of a thick lens \footnote{Pozzi \cite{pozzi_multislice_1989} elegantly showed that one can calculate electron wavefunction propgation through arbitrarily thick electrostatic or magnetostatic optical elements with an analytical multislice formalism. As the lens effect we're interested in will be practically easiest to implement with a thin lens, we only include the thin lens-approximate calculation. However, we follow Pozzi's notation so that an interested reader could complete the thick lens calculation.}. We use the non-relativistic Schrodinger equation for simplicity; low-order relativistic corrections can easily be added into the result.
\begin{equation}
  \nabla^2\psi - \frac{2e}{\hbar i} \mathbf{A}\cdot \boldsymbol{\nabla}\psi + \frac{2m_e}{\hbar^2} \left(eV+E\right)\psi -\frac{e^2}{\hbar^2}A^2\psi = 0
\end{equation}
where $m_e$ is the rest mass of the electron and $E = e V_a$ is the non-relativistic total energy of the electron accelerated by a voltage $V_a$.
If we assume that $\psi$ is separable into
\begin{equation}
  \psi = \psi_0 \chi
\end{equation}
where
\begin{equation}
  \psi_0 = e^{ik_z z}
\end{equation}
we can quickly simplify our Schrodinger equation with a paraxial approximation. If $k_z \approx k = \frac{\sqrt{2 m_e E}}{\hbar}$, we then see that
\begin{equation} \label{eq:parax_condition}
  \nabla^2 \psi_0 + \frac{2 m_e }{\hbar^2}E\psi_0 \approx 0.
\end{equation}
Let's now parse through the terms in the the paraxial Schrodinger equation that results. As
\begin{align}
  \nabla^2 \psi &= \psi_0\nabla_{\bot}^2\chi+2ik_z\psi_0\partial{\chi}{z} - k_z^2\psi_0 \chi+\psi_0\secondpartial{\chi}{z} \nonumber \\
                & \approx \psi_0\nabla_{\bot}^2\chi+2ik_z\psi_0\partial{\chi}{z} - k_z^2\psi_0 \chi
\end{align}
since $\psi_0\secondpartial{\chi}{z} \ll k_z^2\psi_0\chi$, we can use \eqref{eq:parax_condition} and divide out $\psi_0$ to produce the paraxial Schrodinger equation for $\chi$:
\begin{equation} \label{eq:paraxSch}
  \nabla_{\bot}^2\chi + 2ik_z \firstpartial{\chi}{z} -\frac{2e}{i \hbar} \mathbf{A}\cdot \boldsymbol{\nabla}_{\bot}\chi - \frac{2e}{i\hbar}A_z\left(\firstpartial{\chi}{z} + i k_z \chi\right) + \frac{2m_e e}{\hbar^2} V\chi -\frac{e^2}{\hbar^2}A^2\chi = 0
\end{equation}
If we first choose
\begin{equation}
  A_{\phi} =  B_1(z)\frac{\rho}{2}-B_3(z)\frac{\rho^3}{8 b^2}
\end{equation}
we see that we can cancel the higher-order ($\rho^4$ and $\rho^6$) terms in $A^2$ independently with similar terms in $V$ produced by an electrostatic aberration corrector or $A_z$ produced by a multipolar magnetostatic aberration corrector\footnote{Of course, a $\rho^4$ term in $A_z$ produces a $\rho^8$ term in $A^2$, so, as with any multitpolar aberration corrector, we can only push aberrations up to a higher order.} (see section \ref{sect:multipolar_corrector_A}). Let's then solve the Schrodinger equation without these higher-order terms. 
If we furthermore use an orbital angular momentum basis such that
\begin{equation}
  \chi \propto e^{im\phi}
\end{equation}
we see that 
\begin{align}
  \frac{2e}{i\hbar \psi}\mathbf{A}\cdot\boldsymbol{\nabla}\left(\chi\psi\right) &= \frac{2e}{i\hbar}A_{\phi}\frac{1}{\rho}\firstpartial{\chi}{\phi} \nonumber \\
                                                                                &= \frac{2e}{\hbar}\frac{A_{\phi}}{\rho}  m \chi
\end{align}
If we take a thin lens approximation and drop the small $\nabla_{\bot}^2\chi $ term, this resulting equation is a separable first-order differential equation. 
The thin lens-paraxial Schrodinger equation with this vector potential, then, is
\begin{equation}
  2ik_z \firstpartial{\chi}{z} = \frac{e}{\hbar}\left(B_1(z)-B_3(z)\frac{\rho^2}{4b^2}\right)m\chi + \frac{e^2B_1^2(z)}{4\hbar^2}\rho^2\chi 
\end{equation}

Upon integration, we can identify the transfer function of the lens as
\begin{equation} \label{eq:thin_lens_Sch_transfer}
  U_{\textrm{lens}} = \frac{\chi(z\to \infty)}{\chi(z\to-\infty)} =  \exp(i\phi_{\mathrm{Larmor}}) \exp\left(-i\frac{e^2}{8 \hbar^2 k_z}\int_{-\infty}^{\infty} \mathrm{d}z \left[B_1^2(z) - \frac{m \hbar}{e b^2} B_3(z)\right]\rho^2\right)
\end{equation}
where
\begin{align}
  \phi_{\mathrm{Larmor}} &= -\frac{m e}{2\hbar k_z} \int_{-\infty}^{\infty} B_1(z)\mathrm{d}z \\
                         &= - m\sqrt{\frac{e}{8 m_e V_a}} \int_{-\infty}^{\infty} B_1(z)\mathrm{d}z.
\end{align}
We see, by comparison with \eqref{eq:lens_transfer} that
\begin{align}
  \frac{1}{f} &= \frac{e^2 \lambda}{8 \pi \hbar^2 k_z}\int_{-\infty}^{\infty} \mathrm{d}z \left[B_1^2(z) - \frac{m \hbar}{e b^2} B_3(z)\right] \\
  &= \frac{e^2}{8 m_e E}\int_{-\infty}^{\infty} \mathrm{d}z \left[B_1^2(z) - \frac{m \hbar}{e b^2} B_3(z)\right] \\
  &= \frac{e}{8 m_e V_a}\int_{-\infty}^{\infty} \mathrm{d}z \left[B_1^2(z) - \frac{m \hbar}{e b^2} B_3(z)\right]
\end{align}
We thus showed that we can derive the orbital angular momentum-dependent focal length of a thin lens from the paraxial Schrodinger equation. 

\section{relationship between OAM lensing term and spherical aberration}

If, instead of dropping terms above $\rho^2$ in eq. \eqref{eq:paraxSch}, we include up to $\rho^4$, we can calculate the contribution to the spherical aberration coefficient $C_s$ from the OAM dispersion term. Keeping this term in our thin lens-paraxial Schrodinger equation, we see
\begin{equation}
  2ik_z \firstpartial{\chi}{z} = \frac{e}{\hbar}\left(B_1(z)-B_3(z)\frac{\rho^2}{4b^2}\right)m\chi + \frac{e^2B_1^2(z)}{4\hbar^2}\rho^2\chi - \frac{e^2B_1(z)B_3(z)}{8\hbar^2 b^2}\rho^4\chi
\end{equation}
Integrating as in \eqref{eq:thin_lens_Sch_transfer}, our transfer function now includes the term
\begin{equation} \label{eq:sph_transfer}
  U_{\textrm{spherical}} = \exp\left(i\frac{e^2}{16\hbar^2 k_z b^2}\int_{-\infty}^{\infty} \mathrm{d}z \left[B_1(z)B_3(z)\right]\rho^4\right)
\end{equation}
As the aberrations of an electron lens are conventionally expanded in terms of the polar angle of incidence at the back focal plane of the lens $\alpha = \arctan\frac{\rho}{f}$ with a transfer function for the lowest-order spherical aberration \cite{kirkland_optimum_2011},
\begin{equation} \label{eq:conv_sph_transfer}
  U_{\textrm{spherical}} = \exp\left(i\frac{2\pi}{\lambda}\frac{C_3}{4}\alpha^4\right)
\end{equation}
where $C_3$ is the third-order spherical aberration coefficient.
If we rewrite \eqref{eq:sph_transfer} in this form with the approximation that $\rho \approx f \alpha$, we can calculate $C_3$.
\begin{equation} \label{eq:sph_transfer_angle}
  U_{\textrm{spherical}}= \exp\left(i\frac{2\pi}{\lambda}\frac{e^2}{4 \hbar^2 k_z^2 b^2}\int_{-\infty}^{\infty} \mathrm{d}z \left[B_1(z)B_3(z)\right]f^4\frac{\alpha^4}{4}\right)
\end{equation}
By comparison with \eqref{eq:conv_sph_transfer}, we see with some reorganization that in the thin lens approximation, the contribution to $C_3$ from the OAM dispersion term we introduced is
\begin{equation}
  C_3 = \frac{e^2 f^4}{8 m_e E b^2}\int_{-\infty}^{\infty} \mathrm{d}z \left[B_1(z)B_3(z)\right]
\end{equation}
As the OAM dispersion length $b$ must be small to produce significant orbital dispersion, $C_3$ could be prohibitively large under standard transmission electron microscope conditions even with independent reduction of $C_3$ by a multipole corrector (see section \ref{sect:multipolar_corrector_A}). Realization of orbital dispersion that is distinguishable over spherical aberration for 80 to \unit[300]{keV} electrons may demand better aberration correctors than are available today.

\section{Calculation of the OAM-dependent focal length for several field distributions. \label{sect:physical_current}}

The Glaser field, 
\begin{equation} \label{eq:glaser}
  B_{\textrm{G}}(z) = B_0 \left(1+\frac{z^2}{a^2}\right)^{-1}
\end{equation}
is commonly used to model the longitudinal field of a magnetic lens with a longitudinal extent of length $a$. If we choose $B_1 = B_3 = B_{\textrm{G}}$ for our focal length calculation, we see
\begin{equation} \label{eq:glaser_f}
  f_m = \frac{16 m_e E}{e^2 B_0^2 a \pi\left(1- \frac{2 m\hbar}{e B_0 b^2}\right)} \\
\end{equation}

For a loop of wire with radius $R$ and current $I_0$, let's calculate the vector potential for small $\rho \ll R$.
\begin{align}
  \mathbf{A}(\mathbf{r}) &= \frac{\mu_0 I_0 R}{4\pi} \int \mathrm{d}\phi' \frac{\hat{\boldsymbol{\phi}}'}{\sqrt{z^2 +R^2 + \rho^2 + 2R\rho\cos(\phi'-\phi)}} \\
  \mathbf{A}(\mathbf{r}) &\approx \frac{\mu_0 I_0 R}{4\pi \ell(z)} \int \mathrm{d}\phi' \hat{\boldsymbol{\phi}}'\left(1-\frac{1}{2}\frac{\rho^2 + 2R\rho\cos(\phi'-\phi)}{\ell^2(z)}+\frac{3}{8}\left(\frac{\rho^2 + 2R\rho\cos(\phi'-\phi)}{\ell^2(z)}\right)^2- \frac{5}{8}\left(\frac{\rho^2 + 2R\rho\cos(\phi'-\phi)}{\ell^2(z)}\right)^3+\ldots\right)
\end{align}
where $\ell(z) = \sqrt{z^2+R^2}$. If we perform the integral over $\phi'$ and keep terms up to $\rho^3$, we see
\begin{align}
  \mathbf{A}(\mathbf{r}) &= \frac{\mu_0 I_0 R}{4\pi} \left(\frac{\rho R \pi}{\ell^3(z)} - \frac{3}{2} \frac{\rho^3 R\pi}{\ell^5(z)} - \frac{15}{4}\frac{\rho^3 R^3\pi}{\ell^7(z)}+\ldots\right)\hat{\boldsymbol{\phi}}
\end{align}
Using the formalism we developed above, let's calculate the focal length of the lensing behavior produced by this term. First, let's define
\begin{align}
  B_0 = & \frac{\mu_0 I_0}{R} \\
  B_1(z) &= \frac{B_0 R^3}{2 \ell^3(z)} \\
  B_3(z) &= 3 B_0 \left(\frac{R^5}{\ell^5(z)}+\frac{5}{2}\frac{R^7}{\ell^7(z)}\right)
\end{align}
such that can write the vector potential as
\begin{equation}
  \mathbf{A}(\mathbf{r}) \approx \frac{B_1(z)}{2}\rho - \frac{B_3(z)}{8 R^2}\rho^3\hat{\boldsymbol{\phi}}.
\end{equation}
We thus observe that equation \eqref{eq:modelvp} is an accurate physical description up to $\rho^3$. Let's now calculate the focal length of this lens.
We see that, as
\begin{equation}
  \int_{-\infty}^{\infty} B_3(z) \mathrm{d}z = 12 B_0 R
\end{equation}
and as
\begin{equation}
\int_{-\infty}^{\infty} B_1^2(z) \mathrm{d}z = \frac{3\pi}{8} B_0^2 R
\end{equation}
we can write the focal length of this lens, using \eqref{eq:thin_f}, as
\begin{align}
  \frac{1}{f_m} &= \frac{e^2}{8 m_e E} \left(\frac{3\pi}{8}B_0^2 R- \frac{m \hbar}{e R^2} \left(12 B_0 R\right) \right) \\
  f_m &= \frac{64 m_e E}{3 e^2 B_0^2 R \pi \left(1-\frac{32 m \hbar}{e B_0 R^2}\right)}
\end{align}
We see that this result is indentical in form to the Glaser-field result for our simpler model, \eqref{eq:glaser_f}, if the longitudinal extent is set by the radius ($a=R$) and the OAM dispersion length is set by the radius ($b=R$). The focal lengths differ only by constant factors.

\section{Independent correction of aberrations induced by maximizing OAM-dependent focusing\label{sect:multipolar_corrector_A}}

In this section, we show that a multipole corrector has no $A_{\phi}$ component and thus can independently correct aberrations produced by the $A^2$ term of an OAM measurement device without affecting measurement of OAM.

If we represent an $n$-pole magnetic lens as a ring of $n$ solenoids of alternating polarity with the solenoid axis oriented radially, we'll immediately see that the $A_{\phi}$ component of the vector potential would produce a lens with infinite focal length--no lensing effect--in the thin lens approximation. 

First, though, let's write a model for the vector potential of a solenoid oriented along the $z$ axis. As the current of an ideal solenoid is entirely azimuthal and cylindrically symmetric, let's write this vector potential as
\begin{equation}
  \mathbf{A}_{\mathrm{ax}} = A_0(z,\rho)\hat{\boldsymbol{\phi}}.
\end{equation}
If we now rotate this vector potential to point along the $x$ axis, we see
\begin{equation}
  \mathbf{A}_{\mathrm{rad}} = A_0(x,\sqrt{y^2+z^2})\left(\frac{-z}{\sqrt{y^2+z^2}}\hat{\mathbf{y}} + \frac{y}{\sqrt{y^2+z^2}}\hat{\mathbf{z}}\right).
\end{equation}
Lastly, if we define a set of rotated coordinates $(x_m,y_m)$ defined by a rotation angle $\theta_m = \frac{2\pi m}{n}$ where
\begin{align}
  x_m & = x\cos(\theta_m + y\sin(\theta_m) \\
  y_m & = y\cos(\theta_m) - x\sin(\theta_m)
\end{align}
we can now easily write the vector potential of this lens in terms of a sum of solenoidal vector potentials in the rotated coordinates.
\begin{equation}
  \mathbf{A}_{n-\textrm{pole}} = \sum_{m=0}^{n-1} A_0(x_m,\sqrt{y_m^2 + z^2}) \left(\frac{-z}{\sqrt{y_m^2+z^2}}\hat{\mathbf{y}}_m + \frac{y_m}{\sqrt{y_m^2+z^2}}\hat{\mathbf{z}}\right)
\end{equation}
We can immediately see that the $\hat{\mathbf{y}}_m$ component, which includes a non-zero $\hat{\boldsymbol{\phi}}$ term, is odd in $z$ and thus integrates to zero under a calculation of the focal length in the thin lens approximation.
\begin{equation}
  \int_{-\infty}^{\infty} \mathrm{d}z\ \mathbf{A}_{n-\textrm{pole}} \cdot \hat{\boldsymbol{\phi}} = 0 \qquad \Rightarrow \qquad f_{m_{n-\textrm{pole}}} = \infty
\end{equation}

Therefore, a multipolar magnetostatic aberration corrector element has a vanishingly small orbital angular momentum lensing effect, and can safely be used to independently correct higher-order aberrations produced by a round lens without affecting the orbital angular momentum dispersion of that round lens.

\section{Stacked lens OAM measurement device design 1: afocal system, or fixed-separation between lenses \label{sect:stacked1}}
If the OAM dispersion coefficient $\Lambda = \frac{2 \hbar}{eB_0 b^2}$ is small, then the focal length is approximately
\begin{equation}
  f_m = f_0 \left(1 + \Lambda m\right)
\end{equation}
If we set two lenses back-to-back with a distance $2f_0$ in between them with opposite OAM dispersion in each ($\Lambda_1 = -|\Lambda|$; $\Lambda_2 = |\Lambda|$) we produce an afocal system with
\begin{equation}
  M_m = -\left(1+2\Lambda m\right)
\end{equation}
Since an afocal system produces no convergence or divergence--the effective focal length is infinite \cite{malacara-hernandez_handbook_2014}--any combination of afocal systems is also an afocal system; this afocal system is thus easy to stack. In particular, for a stack of $N$ such afocal systems, in the limit of large $N$, the total magnification approaches
\begin{equation} \label{eq:mag_afocal}
  M_m^N = (-1)^N \exp\left(2\Lambda m N\right)
\end{equation}
This set of $N$ afocal systems has one clear advantage: even with arbitrarily small OAM dispersion $\Lambda$, we can easily distinguish between any two orbital angular momentum orders with a sufficiently long stack $N$.

A twenty-element ($N=20$) set of identical afocal systems with $a = \unit[10]{\mu \textrm{m}}$, $b = \unit[100]{\textrm{nm}}$, $B_0 = \unit[2]{\textrm{T}}$ and a resultant $f_0 \approx \unit[60]{\textrm{mm}}$ has an OAM dispersion coefficient $|\Lambda| = 0.066$ and a produces a magnification of an $m=+1$, $\unit[80]{\textrm{keV}}$ electron beam of $|M_{+1}^{20}| = 3.73$; on the other hand, an $m=-1$ beam sees a magnification of $|M_{-1}^{20}| = 0.27$. A superposition of two otherwise-identical $m=+1$ and $m=-1$ modes passed through this device leave with a fourteen-times difference in magnification. The total length of this device is on the order of a couple meters.

\section{Stacked lens OAM measurement device design 2: variable spacing between lenses \label{sect:stacked2}}
\begin{figure}[hb]
  \includegraphics[width=\columnwidth]{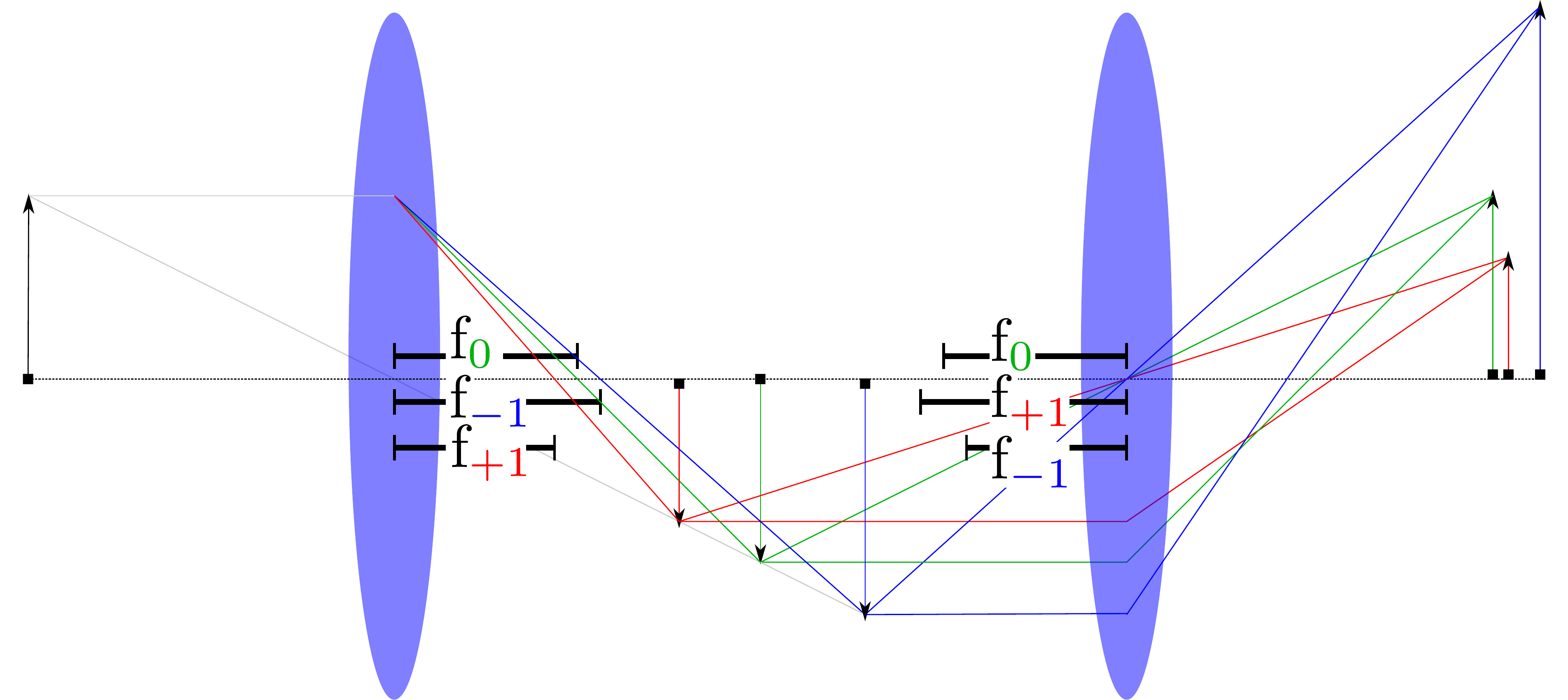}
  \caption{Ray diagram for a combination of two lenses (blue disks) with variable spacing in between and opposite OAM dispersion which combine to produce a strongly OAM-dependent magnification, as given by \eqref{eq:mag_var_spacing} and a weakly OAM-dependent image position.}
  \caption{Cartoon illustration of a lens (blue disk) with OAM-dependent focal length $f_m$ as given in \eqref{eq:glaser_f}. (red) ray diagram for $m=+1$ electrons; (green) ray diagram for $m=0$ electrons; (blue) ray diagram for $m=-1$ electrons. }
\end{figure}

If we place two lenses with opposite OAM dispersion back-to-back with a distance $2\frac{s+1}{s}f_0$ between them, and place an object at a distance $(s+1)f_0$ in front of the first lens, we'll see a focused image at a distance $(s+1)f_0$ behind the second lens with a magnification
\begin{equation} \label{eq:mag_var_spacing}
  M_m =  \frac{1}{1-2(s+1)\Lambda m}
\end{equation}
The result is similar to that for an afocal system with two major differences: the advantage of this system is that larger magnification is produced by a larger spacing, rather than more lenses; the disadvantage is that only one mode can be fully separated from the rest at a time, as if $2(s+1)\Lambda m \approx 1$ so as to maximize magnification of the $m$-OAM components of the beam, then $2(s+1) \Lambda (m+1)$ cannot also be close to $1$ unless $m$ is very large.


\end{widetext}
\bibliographystyle{apsrev4-1}
\bibliography{lensing_spectrometer}{}

\begin{thebibliography}{44}%
\makeatletter
\providecommand \@ifxundefined [1]{%
 \@ifx{#1\undefined}
}%
\providecommand \@ifnum [1]{%
 \ifnum #1\expandafter \@firstoftwo
 \else \expandafter \@secondoftwo
 \fi
}%
\providecommand \@ifx [1]{%
 \ifx #1\expandafter \@firstoftwo
 \else \expandafter \@secondoftwo
 \fi
}%
\providecommand \natexlab [1]{#1}%
\providecommand \enquote  [1]{``#1''}%
\providecommand \bibnamefont  [1]{#1}%
\providecommand \bibfnamefont [1]{#1}%
\providecommand \citenamefont [1]{#1}%
\providecommand \href@noop [0]{\@secondoftwo}%
\providecommand \href [0]{\begingroup \@sanitize@url \@href}%
\providecommand \@href[1]{\@@startlink{#1}\@@href}%
\providecommand \@@href[1]{\endgroup#1\@@endlink}%
\providecommand \@sanitize@url [0]{\catcode `\\12\catcode `\$12\catcode
  `\&12\catcode `\#12\catcode `\^12\catcode `\_12\catcode `\%12\relax}%
\providecommand \@@startlink[1]{}%
\providecommand \@@endlink[0]{}%
\providecommand \url  [0]{\begingroup\@sanitize@url \@url }%
\providecommand \@url [1]{\endgroup\@href {#1}{\urlprefix }}%
\providecommand \urlprefix  [0]{URL }%
\providecommand \Eprint [0]{\href }%
\providecommand \doibase [0]{http://dx.doi.org/}%
\providecommand \selectlanguage [0]{\@gobble}%
\providecommand \bibinfo  [0]{\@secondoftwo}%
\providecommand \bibfield  [0]{\@secondoftwo}%
\providecommand \translation [1]{[#1]}%
\providecommand \BibitemOpen [0]{}%
\providecommand \bibitemStop [0]{}%
\providecommand \bibitemNoStop [0]{.\EOS\space}%
\providecommand \EOS [0]{\spacefactor3000\relax}%
\providecommand \BibitemShut  [1]{\csname bibitem#1\endcsname}%
\let\auto@bib@innerbib\@empty
\bibitem [{\citenamefont {Uchida}\ and\ \citenamefont
  {Tonomura}(2010)}]{uchida_generation_2010}%
  \BibitemOpen
  \bibfield  {author} {\bibinfo {author} {\bibfnamefont {M.}~\bibnamefont
  {Uchida}}\ and\ \bibinfo {author} {\bibfnamefont {A.}~\bibnamefont
  {Tonomura}},\ }\href {\doibase 10.1038/nature08904} {\bibfield  {journal}
  {\bibinfo  {journal} {Nature}\ }\textbf {\bibinfo {volume} {464}},\ \bibinfo
  {pages} {737} (\bibinfo {year} {2010})},\ \bibinfo {note} {00172 Cited by
  0096}\BibitemShut {NoStop}%
\bibitem [{\citenamefont {Verbeeck}\ \emph {et~al.}(2010)\citenamefont
  {Verbeeck}, \citenamefont {Tian},\ and\ \citenamefont
  {Schattschneider}}]{verbeeck_production_2010}%
  \BibitemOpen
  \bibfield  {author} {\bibinfo {author} {\bibfnamefont {J.}~\bibnamefont
  {Verbeeck}}, \bibinfo {author} {\bibfnamefont {H.}~\bibnamefont {Tian}}, \
  and\ \bibinfo {author} {\bibfnamefont {P.}~\bibnamefont {Schattschneider}},\
  }\href {\doibase 10.1038/nature09366} {\bibfield  {journal} {\bibinfo
  {journal} {Nature}\ }\textbf {\bibinfo {volume} {467}},\ \bibinfo {pages}
  {301} (\bibinfo {year} {2010})},\ \bibinfo {note} {00221 Cited by
  0114}\BibitemShut {NoStop}%
\bibitem [{\citenamefont {McMorran}\ \emph {et~al.}(2011)\citenamefont
  {McMorran}, \citenamefont {Agrawal}, \citenamefont {Anderson}, \citenamefont
  {Herzing}, \citenamefont {Lezec}, \citenamefont {McClelland},\ and\
  \citenamefont {Unguris}}]{mcmorran_electron_2011}%
  \BibitemOpen
  \bibfield  {author} {\bibinfo {author} {\bibfnamefont {B.~J.}\ \bibnamefont
  {McMorran}}, \bibinfo {author} {\bibfnamefont {A.}~\bibnamefont {Agrawal}},
  \bibinfo {author} {\bibfnamefont {I.~M.}\ \bibnamefont {Anderson}}, \bibinfo
  {author} {\bibfnamefont {A.~A.}\ \bibnamefont {Herzing}}, \bibinfo {author}
  {\bibfnamefont {H.~J.}\ \bibnamefont {Lezec}}, \bibinfo {author}
  {\bibfnamefont {J.~J.}\ \bibnamefont {McClelland}}, \ and\ \bibinfo {author}
  {\bibfnamefont {J.}~\bibnamefont {Unguris}},\ }\href {\doibase
  10.1126/science.1198804} {\bibfield  {journal} {\bibinfo  {journal}
  {Science}\ }\textbf {\bibinfo {volume} {331}},\ \bibinfo {pages} {192 }
  (\bibinfo {year} {2011})}\BibitemShut {NoStop}%
\bibitem [{\citenamefont {Saitoh}\ \emph {et~al.}(2012)\citenamefont {Saitoh},
  \citenamefont {Hasegawa}, \citenamefont {Tanaka},\ and\ \citenamefont
  {Uchida}}]{saitoh_production_2012}%
  \BibitemOpen
  \bibfield  {author} {\bibinfo {author} {\bibfnamefont {K.}~\bibnamefont
  {Saitoh}}, \bibinfo {author} {\bibfnamefont {Y.}~\bibnamefont {Hasegawa}},
  \bibinfo {author} {\bibfnamefont {N.}~\bibnamefont {Tanaka}}, \ and\ \bibinfo
  {author} {\bibfnamefont {M.}~\bibnamefont {Uchida}},\ }\href {\doibase
  10.1093/jmicro/dfs036} {\bibfield  {journal} {\bibinfo  {journal} {Journal of
  Electron Microscopy}\ }\textbf {\bibinfo {volume} {61}},\ \bibinfo {pages}
  {171} (\bibinfo {year} {2012})}\BibitemShut {NoStop}%
\bibitem [{\citenamefont {Schattschneider}\ \emph
  {et~al.}(2012{\natexlab{a}})\citenamefont {Schattschneider}, \citenamefont
  {St\"{o}ger-Pollach},\ and\ \citenamefont
  {Verbeeck}}]{schattschneider_novel_2012}%
  \BibitemOpen
  \bibfield  {author} {\bibinfo {author} {\bibfnamefont {P.}~\bibnamefont
  {Schattschneider}}, \bibinfo {author} {\bibfnamefont {M.}~\bibnamefont
  {St\"{o}ger-Pollach}}, \ and\ \bibinfo {author} {\bibfnamefont
  {J.}~\bibnamefont {Verbeeck}},\ }\href {\doibase
  10.1103/PhysRevLett.109.084801} {\bibfield  {journal} {\bibinfo  {journal}
  {Physical Review Letters}\ }\textbf {\bibinfo {volume} {109}},\ \bibinfo
  {pages} {084801} (\bibinfo {year} {2012}{\natexlab{a}})},\ \bibinfo {note}
  {00035}\BibitemShut {NoStop}%
\bibitem [{\citenamefont {Blackburn}\ and\ \citenamefont
  {Loudon}(2014)}]{blackburn_vortex_2014}%
  \BibitemOpen
  \bibfield  {author} {\bibinfo {author} {\bibfnamefont {A.~M.}\ \bibnamefont
  {Blackburn}}\ and\ \bibinfo {author} {\bibfnamefont {J.~C.}\ \bibnamefont
  {Loudon}},\ }\href {\doibase 10.1016/j.ultramic.2013.08.009} {\bibfield
  {journal} {\bibinfo  {journal} {Ultramicroscopy}\ }\textbf {\bibinfo {volume}
  {136}},\ \bibinfo {pages} {127} (\bibinfo {year} {2014})},\ \bibinfo {note}
  {00013}\BibitemShut {NoStop}%
\bibitem [{\citenamefont {B\'{e}ch\'{e}}\ \emph {et~al.}(2014)\citenamefont
  {B\'{e}ch\'{e}}, \citenamefont {Van~Boxem}, \citenamefont {Van~Tendeloo},\
  and\ \citenamefont {Verbeeck}}]{beche_magnetic_2014}%
  \BibitemOpen
  \bibfield  {author} {\bibinfo {author} {\bibfnamefont {A.}~\bibnamefont
  {B\'{e}ch\'{e}}}, \bibinfo {author} {\bibfnamefont {R.}~\bibnamefont
  {Van~Boxem}}, \bibinfo {author} {\bibfnamefont {G.}~\bibnamefont
  {Van~Tendeloo}}, \ and\ \bibinfo {author} {\bibfnamefont {J.}~\bibnamefont
  {Verbeeck}},\ }\href {\doibase 10.1038/nphys2816} {\bibfield  {journal}
  {\bibinfo  {journal} {Nature Physics}\ }\textbf {\bibinfo {volume} {10}},\
  \bibinfo {pages} {26} (\bibinfo {year} {2014})},\ \bibinfo {note}
  {00034}\BibitemShut {NoStop}%
\bibitem [{\citenamefont {Harvey}\ \emph {et~al.}(2014)\citenamefont {Harvey},
  \citenamefont {Pierce}, \citenamefont {Agrawal}, \citenamefont {Ercius},
  \citenamefont {Linck},\ and\ \citenamefont
  {McMorran}}]{harvey_efficient_2014}%
  \BibitemOpen
  \bibfield  {author} {\bibinfo {author} {\bibfnamefont {T.~R.}\ \bibnamefont
  {Harvey}}, \bibinfo {author} {\bibfnamefont {J.~S.}\ \bibnamefont {Pierce}},
  \bibinfo {author} {\bibfnamefont {A.~K.}\ \bibnamefont {Agrawal}}, \bibinfo
  {author} {\bibfnamefont {P.}~\bibnamefont {Ercius}}, \bibinfo {author}
  {\bibfnamefont {M.}~\bibnamefont {Linck}}, \ and\ \bibinfo {author}
  {\bibfnamefont {B.~J.}\ \bibnamefont {McMorran}},\ }\href {\doibase
  10.1088/1367-2630/16/9/093039} {\bibfield  {journal} {\bibinfo  {journal}
  {New Journal of Physics}\ }\textbf {\bibinfo {volume} {16}},\ \bibinfo
  {pages} {093039} (\bibinfo {year} {2014})},\ \bibinfo {note}
  {00010}\BibitemShut {NoStop}%
\bibitem [{\citenamefont {Grillo}\ \emph {et~al.}(2014)\citenamefont {Grillo},
  \citenamefont {Gazzadi}, \citenamefont {Karimi}, \citenamefont {Mafakheri},
  \citenamefont {Boyd},\ and\ \citenamefont {Frabboni}}]{grillo_highly_2014}%
  \BibitemOpen
  \bibfield  {author} {\bibinfo {author} {\bibfnamefont {V.}~\bibnamefont
  {Grillo}}, \bibinfo {author} {\bibfnamefont {G.~C.}\ \bibnamefont {Gazzadi}},
  \bibinfo {author} {\bibfnamefont {E.}~\bibnamefont {Karimi}}, \bibinfo
  {author} {\bibfnamefont {E.}~\bibnamefont {Mafakheri}}, \bibinfo {author}
  {\bibfnamefont {R.~W.}\ \bibnamefont {Boyd}}, \ and\ \bibinfo {author}
  {\bibfnamefont {S.}~\bibnamefont {Frabboni}},\ }\href {\doibase
  10.1063/1.4863564} {\bibfield  {journal} {\bibinfo  {journal} {Applied
  Physics Letters}\ }\textbf {\bibinfo {volume} {104}},\ \bibinfo {pages}
  {043109} (\bibinfo {year} {2014})}\BibitemShut {NoStop}%
\bibitem [{\citenamefont {Shiloh}\ \emph {et~al.}(2014)\citenamefont {Shiloh},
  \citenamefont {Lereah}, \citenamefont {Lilach},\ and\ \citenamefont
  {Arie}}]{shiloh_sculpturing_2014}%
  \BibitemOpen
  \bibfield  {author} {\bibinfo {author} {\bibfnamefont {R.}~\bibnamefont
  {Shiloh}}, \bibinfo {author} {\bibfnamefont {Y.}~\bibnamefont {Lereah}},
  \bibinfo {author} {\bibfnamefont {Y.}~\bibnamefont {Lilach}}, \ and\ \bibinfo
  {author} {\bibfnamefont {A.}~\bibnamefont {Arie}},\ }\href {\doibase
  10.1016/j.ultramic.2014.04.007} {\bibfield  {journal} {\bibinfo  {journal}
  {Ultramicroscopy}\ }\textbf {\bibinfo {volume} {144}},\ \bibinfo {pages} {26}
  (\bibinfo {year} {2014})}\BibitemShut {NoStop}%
\bibitem [{\citenamefont {B\'{e}ch\'{e}}\ \emph {et~al.}(2016)\citenamefont
  {B\'{e}ch\'{e}}, \citenamefont {Juchtmans},\ and\ \citenamefont
  {Verbeeck}}]{beche_efficient_????}%
  \BibitemOpen
  \bibfield  {author} {\bibinfo {author} {\bibfnamefont {A.}~\bibnamefont
  {B\'{e}ch\'{e}}}, \bibinfo {author} {\bibfnamefont {R.}~\bibnamefont
  {Juchtmans}}, \ and\ \bibinfo {author} {\bibfnamefont {J.}~\bibnamefont
  {Verbeeck}},\ }\href {\doibase 10.1016/j.ultramic.2016.05.006} {\bibfield
  {journal} {\bibinfo  {journal} {Ultramicroscopy}\ } (\bibinfo {year}
  {2016}),\ 10.1016/j.ultramic.2016.05.006}\BibitemShut {NoStop}%
\bibitem [{\citenamefont {Schattschneider}\ \emph
  {et~al.}(2012{\natexlab{b}})\citenamefont {Schattschneider}, \citenamefont
  {Schaffer}, \citenamefont {Ennen},\ and\ \citenamefont
  {Verbeeck}}]{schattschneider_mapping_2012}%
  \BibitemOpen
  \bibfield  {author} {\bibinfo {author} {\bibfnamefont {P.}~\bibnamefont
  {Schattschneider}}, \bibinfo {author} {\bibfnamefont {B.}~\bibnamefont
  {Schaffer}}, \bibinfo {author} {\bibfnamefont {I.}~\bibnamefont {Ennen}}, \
  and\ \bibinfo {author} {\bibfnamefont {J.}~\bibnamefont {Verbeeck}},\ }\href
  {\doibase 10.1103/PhysRevB.85.134422} {\bibfield  {journal} {\bibinfo
  {journal} {Physical Review B}\ }\textbf {\bibinfo {volume} {85}},\ \bibinfo
  {pages} {134422} (\bibinfo {year} {2012}{\natexlab{b}})}\BibitemShut
  {NoStop}%
\bibitem [{\citenamefont {Asenjo-Garcia}\ and\ \citenamefont {Garc\'{i}a~de
  Abajo}(2014)}]{asenjo-garcia_dichroism_2014}%
  \BibitemOpen
  \bibfield  {author} {\bibinfo {author} {\bibfnamefont {A.}~\bibnamefont
  {Asenjo-Garcia}}\ and\ \bibinfo {author} {\bibfnamefont {F.}~\bibnamefont
  {Garc\'{i}a~de Abajo}},\ }\href {\doibase 10.1103/PhysRevLett.113.066102}
  {\bibfield  {journal} {\bibinfo  {journal} {Physical Review Letters}\
  }\textbf {\bibinfo {volume} {113}},\ \bibinfo {pages} {066102} (\bibinfo
  {year} {2014})},\ \bibinfo {note} {00000}\BibitemShut {NoStop}%
\bibitem [{\citenamefont {Harvey}\ \emph {et~al.}(2015)\citenamefont {Harvey},
  \citenamefont {Pierce}, \citenamefont {Chess},\ and\ \citenamefont
  {McMorran}}]{harvey_demonstration_2015}%
  \BibitemOpen
  \bibfield  {author} {\bibinfo {author} {\bibfnamefont {T.~R.}\ \bibnamefont
  {Harvey}}, \bibinfo {author} {\bibfnamefont {J.~S.}\ \bibnamefont {Pierce}},
  \bibinfo {author} {\bibfnamefont {J.~J.}\ \bibnamefont {Chess}}, \ and\
  \bibinfo {author} {\bibfnamefont {B.~J.}\ \bibnamefont {McMorran}},\ }\href
  {http://arxiv.org/abs/1507.01810} {\bibfield  {journal} {\bibinfo  {journal}
  {arXiv:1507.01810 [cond-mat, physics]}\ } (\bibinfo {year} {2015})},\
  \bibinfo {note} {00001 arXiv: 1507.01810}\BibitemShut {NoStop}%
\bibitem [{Note1()}]{Note1}%
  \BibitemOpen
  \bibinfo {note} {Simulations \cite {rusz_achieving_2014} suggest that a small
  dichroism effect does exist when one measures only the probability density of
  the final state--and therefore traces out OAM in the final
  state.}\BibitemShut {Stop}%
\bibitem [{\citenamefont {Macek}\ \emph {et~al.}(2010)\citenamefont {Macek},
  \citenamefont {Sternberg}, \citenamefont {Ovchinnikov},\ and\ \citenamefont
  {Briggs}}]{macek_theory_2010}%
  \BibitemOpen
  \bibfield  {author} {\bibinfo {author} {\bibfnamefont {J.~H.}\ \bibnamefont
  {Macek}}, \bibinfo {author} {\bibfnamefont {J.~B.}\ \bibnamefont
  {Sternberg}}, \bibinfo {author} {\bibfnamefont {S.~Y.}\ \bibnamefont
  {Ovchinnikov}}, \ and\ \bibinfo {author} {\bibfnamefont {J.~S.}\ \bibnamefont
  {Briggs}},\ }\href {\doibase 10.1103/PhysRevLett.104.033201} {\bibfield
  {journal} {\bibinfo  {journal} {Physical Review Letters}\ }\textbf {\bibinfo
  {volume} {104}},\ \bibinfo {pages} {033201} (\bibinfo {year} {2010})},\
  \bibinfo {note} {00000}\BibitemShut {NoStop}%
\bibitem [{\citenamefont {Ward}\ and\ \citenamefont
  {Macek}(2014)}]{ward_effect_2014}%
  \BibitemOpen
  \bibfield  {author} {\bibinfo {author} {\bibfnamefont {S.~J.}\ \bibnamefont
  {Ward}}\ and\ \bibinfo {author} {\bibfnamefont {J.~H.}\ \bibnamefont
  {Macek}},\ }\href {\doibase 10.1103/PhysRevA.90.062709} {\bibfield  {journal}
  {\bibinfo  {journal} {Physical Review A}\ }\textbf {\bibinfo {volume} {90}},\
  \bibinfo {pages} {062709} (\bibinfo {year} {2014})}\BibitemShut {NoStop}%
\bibitem [{\citenamefont {Ngoko~Djiokap}\ \emph {et~al.}(2015)\citenamefont
  {Ngoko~Djiokap}, \citenamefont {Hu}, \citenamefont {Madsen}, \citenamefont
  {Manakov}, \citenamefont {Meremianin},\ and\ \citenamefont
  {Starace}}]{ngoko_djiokap_electron_2015}%
  \BibitemOpen
  \bibfield  {author} {\bibinfo {author} {\bibfnamefont {J.}~\bibnamefont
  {Ngoko~Djiokap}}, \bibinfo {author} {\bibfnamefont {S.}~\bibnamefont {Hu}},
  \bibinfo {author} {\bibfnamefont {L.}~\bibnamefont {Madsen}}, \bibinfo
  {author} {\bibfnamefont {N.}~\bibnamefont {Manakov}}, \bibinfo {author}
  {\bibfnamefont {A.}~\bibnamefont {Meremianin}}, \ and\ \bibinfo {author}
  {\bibfnamefont {A.~F.}\ \bibnamefont {Starace}},\ }\href {\doibase
  10.1103/PhysRevLett.115.113004} {\bibfield  {journal} {\bibinfo  {journal}
  {Physical Review Letters}\ }\textbf {\bibinfo {volume} {115}},\ \bibinfo
  {pages} {113004} (\bibinfo {year} {2015})}\BibitemShut {NoStop}%
\bibitem [{\citenamefont {G\'{e}neaux}\ \emph {et~al.}(2015)\citenamefont
  {G\'{e}neaux}, \citenamefont {Camper}, \citenamefont {Auguste}, \citenamefont
  {Gobert}, \citenamefont {Caillat}, \citenamefont {Tae\"{i}b},\ and\
  \citenamefont {Ruchon}}]{geneaux_attosecond_2015}%
  \BibitemOpen
  \bibfield  {author} {\bibinfo {author} {\bibfnamefont {R.}~\bibnamefont
  {G\'{e}neaux}}, \bibinfo {author} {\bibfnamefont {A.}~\bibnamefont {Camper}},
  \bibinfo {author} {\bibfnamefont {T.}~\bibnamefont {Auguste}}, \bibinfo
  {author} {\bibfnamefont {O.}~\bibnamefont {Gobert}}, \bibinfo {author}
  {\bibfnamefont {J.}~\bibnamefont {Caillat}}, \bibinfo {author} {\bibfnamefont
  {R.}~\bibnamefont {Tae\"{i}b}}, \ and\ \bibinfo {author} {\bibfnamefont
  {T.}~\bibnamefont {Ruchon}},\ }\href {http://arxiv.org/abs/1509.07396}
  {\bibfield  {journal} {\bibinfo  {journal} {arXiv:1509.07396 [physics]}\ }
  (\bibinfo {year} {2015})},\ \bibinfo {note} {arXiv: 1509.07396}\BibitemShut
  {NoStop}%
\bibitem [{\citenamefont {Serbo}\ \emph {et~al.}(2015)\citenamefont {Serbo},
  \citenamefont {Ivanov}, \citenamefont {Fritzsche}, \citenamefont {Seipt},\
  and\ \citenamefont {Surzhykov}}]{serbo_scattering_2015}%
  \BibitemOpen
  \bibfield  {author} {\bibinfo {author} {\bibfnamefont {V.}~\bibnamefont
  {Serbo}}, \bibinfo {author} {\bibfnamefont {I.~P.}\ \bibnamefont {Ivanov}},
  \bibinfo {author} {\bibfnamefont {S.}~\bibnamefont {Fritzsche}}, \bibinfo
  {author} {\bibfnamefont {D.}~\bibnamefont {Seipt}}, \ and\ \bibinfo {author}
  {\bibfnamefont {A.}~\bibnamefont {Surzhykov}},\ }\href
  {http://arxiv.org/abs/1505.02587} {\bibfield  {journal} {\bibinfo  {journal}
  {arXiv:1505.02587 [hep-ph, physics:physics]}\ } (\bibinfo {year} {2015})},\
  \bibinfo {note} {arXiv: 1505.02587}\BibitemShut {NoStop}%
\bibitem [{\citenamefont {Takahashi}\ and\ \citenamefont
  {Nagaosa}(2015)}]{takahashi_berry_2015}%
  \BibitemOpen
  \bibfield  {author} {\bibinfo {author} {\bibfnamefont {R.}~\bibnamefont
  {Takahashi}}\ and\ \bibinfo {author} {\bibfnamefont {N.}~\bibnamefont
  {Nagaosa}},\ }\href {\doibase 10.1103/PhysRevB.91.245133} {\bibfield
  {journal} {\bibinfo  {journal} {Physical Review B}\ }\textbf {\bibinfo
  {volume} {91}},\ \bibinfo {pages} {245133} (\bibinfo {year} {2015})},\
  \bibinfo {note} {00000}\BibitemShut {NoStop}%
\bibitem [{\citenamefont {Sch\"{u}ler}\ \emph {et~al.}(2015)\citenamefont
  {Sch\"{u}ler}, \citenamefont {Berakdar},\ and\ \citenamefont
  {Pavlyukh}}]{schuler_disentangling_2015}%
  \BibitemOpen
  \bibfield  {author} {\bibinfo {author} {\bibfnamefont {M.}~\bibnamefont
  {Sch\"{u}ler}}, \bibinfo {author} {\bibfnamefont {J.}~\bibnamefont
  {Berakdar}}, \ and\ \bibinfo {author} {\bibfnamefont {Y.}~\bibnamefont
  {Pavlyukh}},\ }\href {\doibase 10.1103/PhysRevA.92.021403} {\bibfield
  {journal} {\bibinfo  {journal} {Physical Review A}\ }\textbf {\bibinfo
  {volume} {92}} (\bibinfo {year} {2015}),\ 10.1103/PhysRevA.92.021403},\
  \bibinfo {note} {arXiv: 1507.04173}\BibitemShut {NoStop}%
\bibitem [{\citenamefont {Karimi}\ \emph {et~al.}(2012)\citenamefont {Karimi},
  \citenamefont {Marrucci}, \citenamefont {Grillo},\ and\ \citenamefont
  {Santamato}}]{karimi_spin--orbital_2012}%
  \BibitemOpen
  \bibfield  {author} {\bibinfo {author} {\bibfnamefont {E.}~\bibnamefont
  {Karimi}}, \bibinfo {author} {\bibfnamefont {L.}~\bibnamefont {Marrucci}},
  \bibinfo {author} {\bibfnamefont {V.}~\bibnamefont {Grillo}}, \ and\ \bibinfo
  {author} {\bibfnamefont {E.}~\bibnamefont {Santamato}},\ }\href {\doibase
  10.1103/PhysRevLett.108.044801} {\bibfield  {journal} {\bibinfo  {journal}
  {Physical Review Letters}\ }\textbf {\bibinfo {volume} {108}},\ \bibinfo
  {pages} {044801} (\bibinfo {year} {2012})}\BibitemShut {NoStop}%
\bibitem [{\citenamefont {Ivanov}(2012)}]{ivanov_measuring_2012}%
  \BibitemOpen
  \bibfield  {author} {\bibinfo {author} {\bibfnamefont {I.~P.}\ \bibnamefont
  {Ivanov}},\ }\href {\doibase 10.1103/PhysRevD.85.076001} {\bibfield
  {journal} {\bibinfo  {journal} {Physical Review D}\ }\textbf {\bibinfo
  {volume} {85}},\ \bibinfo {pages} {076001} (\bibinfo {year} {2012})},\
  \bibinfo {note} {00010}\BibitemShut {NoStop}%
\bibitem [{\citenamefont {Guzzinati}\ \emph {et~al.}(2014)\citenamefont
  {Guzzinati}, \citenamefont {Clark}, \citenamefont {B\'{e}ch\'{e}},\ and\
  \citenamefont {Verbeeck}}]{guzzinati_measuring_2014}%
  \BibitemOpen
  \bibfield  {author} {\bibinfo {author} {\bibfnamefont {G.}~\bibnamefont
  {Guzzinati}}, \bibinfo {author} {\bibfnamefont {L.}~\bibnamefont {Clark}},
  \bibinfo {author} {\bibfnamefont {A.}~\bibnamefont {B\'{e}ch\'{e}}}, \ and\
  \bibinfo {author} {\bibfnamefont {J.}~\bibnamefont {Verbeeck}},\ }\href
  {\doibase 10.1103/PhysRevA.89.025803} {\bibfield  {journal} {\bibinfo
  {journal} {Physical Review A}\ }\textbf {\bibinfo {volume} {89}},\ \bibinfo
  {pages} {025803} (\bibinfo {year} {2014})},\ \bibinfo {note}
  {00006}\BibitemShut {NoStop}%
\bibitem [{\citenamefont {Clark}\ \emph {et~al.}(2014)\citenamefont {Clark},
  \citenamefont {B\'{e}ch\'{e}}, \citenamefont {Guzzinati},\ and\ \citenamefont
  {Verbeeck}}]{clark_quantitative_2014}%
  \BibitemOpen
  \bibfield  {author} {\bibinfo {author} {\bibfnamefont {L.}~\bibnamefont
  {Clark}}, \bibinfo {author} {\bibfnamefont {A.}~\bibnamefont
  {B\'{e}ch\'{e}}}, \bibinfo {author} {\bibfnamefont {G.}~\bibnamefont
  {Guzzinati}}, \ and\ \bibinfo {author} {\bibfnamefont {J.}~\bibnamefont
  {Verbeeck}},\ }\href {http://arxiv.org/abs/1403.4398} {\bibfield  {journal}
  {\bibinfo  {journal} {arXiv:1403.4398 [physics]}\ } (\bibinfo {year}
  {2014})}\BibitemShut {NoStop}%
\bibitem [{\citenamefont {Shiloh}\ \emph {et~al.}(2015)\citenamefont {Shiloh},
  \citenamefont {Tsur}, \citenamefont {Remez}, \citenamefont {Lereah},
  \citenamefont {Malomed}, \citenamefont {Shvedov}, \citenamefont {Hnatovsky},
  \citenamefont {Krolikowski},\ and\ \citenamefont
  {Arie}}]{shiloh_unveiling_2015}%
  \BibitemOpen
  \bibfield  {author} {\bibinfo {author} {\bibfnamefont {R.}~\bibnamefont
  {Shiloh}}, \bibinfo {author} {\bibfnamefont {Y.}~\bibnamefont {Tsur}},
  \bibinfo {author} {\bibfnamefont {R.}~\bibnamefont {Remez}}, \bibinfo
  {author} {\bibfnamefont {Y.}~\bibnamefont {Lereah}}, \bibinfo {author}
  {\bibfnamefont {B.~A.}\ \bibnamefont {Malomed}}, \bibinfo {author}
  {\bibfnamefont {V.}~\bibnamefont {Shvedov}}, \bibinfo {author} {\bibfnamefont
  {C.}~\bibnamefont {Hnatovsky}}, \bibinfo {author} {\bibfnamefont
  {W.}~\bibnamefont {Krolikowski}}, \ and\ \bibinfo {author} {\bibfnamefont
  {A.}~\bibnamefont {Arie}},\ }\href {\doibase 10.1103/PhysRevLett.114.096102}
  {\bibfield  {journal} {\bibinfo  {journal} {Physical Review Letters}\
  }\textbf {\bibinfo {volume} {114}},\ \bibinfo {pages} {096102} (\bibinfo
  {year} {2015})},\ \bibinfo {note} {00004}\BibitemShut {NoStop}%
\bibitem [{\citenamefont {Shutova}\ \emph {et~al.}(2016)\citenamefont
  {Shutova}, \citenamefont {Zhdanova},\ and\ \citenamefont
  {Sokolov}}]{shutova_measurement_2016}%
  \BibitemOpen
  \bibfield  {author} {\bibinfo {author} {\bibfnamefont {M.}~\bibnamefont
  {Shutova}}, \bibinfo {author} {\bibfnamefont {A.}~\bibnamefont {Zhdanova}}, \
  and\ \bibinfo {author} {\bibfnamefont {A.}~\bibnamefont {Sokolov}},\ }in\
  \href {http://meetings.aps.org/Meeting/DAMOP16/Session/Q1.66} {\emph
  {\bibinfo {booktitle} {Bulletin of the {American} {Physical} {Society}}}}\
  (\bibinfo  {publisher} {American Physical Society},\ \bibinfo {year}
  {2016})\BibitemShut {NoStop}%
\bibitem [{\citenamefont {Yahn}\ \emph {et~al.}(2013)\citenamefont {Yahn},
  \citenamefont {Pierce}, \citenamefont {Harvey},\ and\ \citenamefont
  {McMorran}}]{yahn_addition_2013}%
  \BibitemOpen
  \bibfield  {author} {\bibinfo {author} {\bibfnamefont {T.}~\bibnamefont
  {Yahn}}, \bibinfo {author} {\bibfnamefont {J.}~\bibnamefont {Pierce}},
  \bibinfo {author} {\bibfnamefont {T.}~\bibnamefont {Harvey}}, \ and\ \bibinfo
  {author} {\bibfnamefont {B.}~\bibnamefont {McMorran}},\ }\href {\doibase
  10.1017/S1431927613007824} {\bibfield  {journal} {\bibinfo  {journal}
  {Microscopy and Microanalysis}\ }\textbf {\bibinfo {volume} {19}},\ \bibinfo
  {pages} {1166} (\bibinfo {year} {2013})},\ \bibinfo {note}
  {00001}\BibitemShut {NoStop}%
\bibitem [{\citenamefont {Saitoh}\ \emph {et~al.}(2013)\citenamefont {Saitoh},
  \citenamefont {Hasegawa}, \citenamefont {Hirakawa}, \citenamefont {Tanaka},\
  and\ \citenamefont {Uchida}}]{saitoh_measuring_2013}%
  \BibitemOpen
  \bibfield  {author} {\bibinfo {author} {\bibfnamefont {K.}~\bibnamefont
  {Saitoh}}, \bibinfo {author} {\bibfnamefont {Y.}~\bibnamefont {Hasegawa}},
  \bibinfo {author} {\bibfnamefont {K.}~\bibnamefont {Hirakawa}}, \bibinfo
  {author} {\bibfnamefont {N.}~\bibnamefont {Tanaka}}, \ and\ \bibinfo {author}
  {\bibfnamefont {M.}~\bibnamefont {Uchida}},\ }\href {\doibase
  10.1103/PhysRevLett.111.074801} {\bibfield  {journal} {\bibinfo  {journal}
  {Physical Review Letters}\ }\textbf {\bibinfo {volume} {111}},\ \bibinfo
  {pages} {074801} (\bibinfo {year} {2013})}\BibitemShut {NoStop}%
\bibitem [{\citenamefont {Gerlach}\ and\ \citenamefont
  {Stern}(1922)}]{gerlach_experimentelle_1922}%
  \BibitemOpen
  \bibfield  {author} {\bibinfo {author} {\bibfnamefont {W.}~\bibnamefont
  {Gerlach}}\ and\ \bibinfo {author} {\bibfnamefont {O.}~\bibnamefont
  {Stern}},\ }\href {\doibase 10.1007/BF01326983} {\bibfield  {journal}
  {\bibinfo  {journal} {Zeitschrift f\"{u}r Physik}\ }\textbf {\bibinfo
  {volume} {9}},\ \bibinfo {pages} {349} (\bibinfo {year} {1922})},\ \bibinfo
  {note} {00419}\BibitemShut {NoStop}%
\bibitem [{\citenamefont {Noether}(1971)}]{noether_invariant_1971}%
  \BibitemOpen
  \bibfield  {author} {\bibinfo {author} {\bibfnamefont {E.}~\bibnamefont
  {Noether}},\ }\href {\doibase 10.1080/00411457108231446} {\bibfield
  {journal} {\bibinfo  {journal} {Transport Theory and Statistical Physics}\
  }\textbf {\bibinfo {volume} {1}},\ \bibinfo {pages} {186} (\bibinfo {year}
  {1971})}\BibitemShut {NoStop}%
\bibitem [{\citenamefont {Batelaan}\ \emph {et~al.}(1997)\citenamefont
  {Batelaan}, \citenamefont {Gay},\ and\ \citenamefont
  {Schwendiman}}]{batelaan_stern-gerlach_1997}%
  \BibitemOpen
  \bibfield  {author} {\bibinfo {author} {\bibfnamefont {H.}~\bibnamefont
  {Batelaan}}, \bibinfo {author} {\bibfnamefont {T.~J.}\ \bibnamefont {Gay}}, \
  and\ \bibinfo {author} {\bibfnamefont {J.~J.}\ \bibnamefont {Schwendiman}},\
  }\href {\doibase 10.1103/PhysRevLett.79.4517} {\bibfield  {journal} {\bibinfo
   {journal} {Physical Review Letters}\ }\textbf {\bibinfo {volume} {79}},\
  \bibinfo {pages} {4517} (\bibinfo {year} {1997})}\BibitemShut {NoStop}%
\bibitem [{\citenamefont {Reimer}\ and\ \citenamefont
  {Kohl}(2008)}]{reimer_transmission_2008}%
  \BibitemOpen
  \bibfield  {author} {\bibinfo {author} {\bibfnamefont {L.}~\bibnamefont
  {Reimer}}\ and\ \bibinfo {author} {\bibfnamefont {H.}~\bibnamefont {Kohl}},\
  }\href@noop {} {\emph {\bibinfo {title} {Transmission {Electron}
  {Microscopy}: {Physics} {If} {Image} {Formation}}}}\ (\bibinfo  {publisher}
  {Springer},\ \bibinfo {year} {2008})\ \bibinfo {note} {00020}\BibitemShut
  {NoStop}%
\bibitem [{\citenamefont {Grillo}\ \emph {et~al.}(2013)\citenamefont {Grillo},
  \citenamefont {Marrucci}, \citenamefont {Karimi}, \citenamefont {Zanella},\
  and\ \citenamefont {Santamato}}]{grillo_quantum_2013}%
  \BibitemOpen
  \bibfield  {author} {\bibinfo {author} {\bibfnamefont {V.}~\bibnamefont
  {Grillo}}, \bibinfo {author} {\bibfnamefont {L.}~\bibnamefont {Marrucci}},
  \bibinfo {author} {\bibfnamefont {E.}~\bibnamefont {Karimi}}, \bibinfo
  {author} {\bibfnamefont {R.}~\bibnamefont {Zanella}}, \ and\ \bibinfo
  {author} {\bibfnamefont {E.}~\bibnamefont {Santamato}},\ }\href {\doibase
  10.1088/1367-2630/15/9/093026} {\bibfield  {journal} {\bibinfo  {journal}
  {New Journal of Physics}\ }\textbf {\bibinfo {volume} {15}},\ \bibinfo
  {pages} {093026} (\bibinfo {year} {2013})}\BibitemShut {NoStop}%
\bibitem [{Note2()}]{Note2}%
  \BibitemOpen
  \bibinfo {note} {The non-unitarity of the lowest-order approximation to the
  transformation induced by $L_z$ terms produces a small, unphysical loss of
  probability density with each slice.}\BibitemShut {Stop}%
\bibitem [{\citenamefont {Feist}\ \emph {et~al.}(2015)\citenamefont {Feist},
  \citenamefont {Echternkamp}, \citenamefont {Schauss}, \citenamefont
  {Yalunin}, \citenamefont {Sch\"{a}fer},\ and\ \citenamefont
  {Ropers}}]{feist_quantum_2015}%
  \BibitemOpen
  \bibfield  {author} {\bibinfo {author} {\bibfnamefont {A.}~\bibnamefont
  {Feist}}, \bibinfo {author} {\bibfnamefont {K.~E.}\ \bibnamefont
  {Echternkamp}}, \bibinfo {author} {\bibfnamefont {J.}~\bibnamefont
  {Schauss}}, \bibinfo {author} {\bibfnamefont {S.~V.}\ \bibnamefont
  {Yalunin}}, \bibinfo {author} {\bibfnamefont {S.}~\bibnamefont
  {Sch\"{a}fer}}, \ and\ \bibinfo {author} {\bibfnamefont {C.}~\bibnamefont
  {Ropers}},\ }\href {\doibase 10.1038/nature14463} {\bibfield  {journal}
  {\bibinfo  {journal} {Nature}\ }\textbf {\bibinfo {volume} {521}},\ \bibinfo
  {pages} {200} (\bibinfo {year} {2015})},\ \bibinfo {note} {00029}\BibitemShut
  {NoStop}%
\bibitem [{\citenamefont {Marai}\ and\ \citenamefont
  {Mulvey}(1976)}]{marai_scherzers_1976}%
  \BibitemOpen
  \bibfield  {author} {\bibinfo {author} {\bibfnamefont {F.~Z.}\ \bibnamefont
  {Marai}}\ and\ \bibinfo {author} {\bibfnamefont {T.}~\bibnamefont {Mulvey}},\
  }\href {\doibase 10.1016/S0304-3991(76)91130-X} {\bibfield  {journal}
  {\bibinfo  {journal} {Ultramicroscopy}\ }\textbf {\bibinfo {volume} {2}},\
  \bibinfo {pages} {187} (\bibinfo {year} {1976})}\BibitemShut {NoStop}%
\bibitem [{Note3()}]{Note3}%
  \BibitemOpen
  \bibinfo {note} {Pozzi \cite {pozzi_multislice_1989} elegantly showed that
  one can calculate electron wavefunction propgation through arbitrarily thick
  electrostatic or magnetostatic optical elements with an analytical multislice
  formalism. As the lens effect we're interested in will be practically easiest
  to implement with a thin lens, we only include the thin lens-approximate
  calculation. However, we follow Pozzi's notation so that an interested reader
  could complete the thick lens calculation.}\BibitemShut {Stop}%
\bibitem [{Note4()}]{Note4}%
  \BibitemOpen
  \bibinfo {note} {Of course, a $\rho ^4$ term in $A_z$ produces a $\rho ^8$
  term in $A^2$, so, as with any multitpolar aberration corrector, we can only
  push aberrations up to a higher order.}\BibitemShut {Stop}%
\bibitem [{\citenamefont {Kirkland}(2011)}]{kirkland_optimum_2011}%
  \BibitemOpen
  \bibfield  {author} {\bibinfo {author} {\bibfnamefont {E.~J.}\ \bibnamefont
  {Kirkland}},\ }\href {\doibase 10.1016/j.ultramic.2011.09.002} {\bibfield
  {journal} {\bibinfo  {journal} {Ultramicroscopy}\ }\textbf {\bibinfo {volume}
  {111}},\ \bibinfo {pages} {1523} (\bibinfo {year} {2011})},\ \bibinfo {note}
  {00012}\BibitemShut {NoStop}%
\bibitem [{\citenamefont {Malacara-Hern\'{a}ndez}\ and\ \citenamefont
  {Malacara-Hern\'{a}ndez}(2014)}]{malacara-hernandez_handbook_2014}%
  \BibitemOpen
  \bibfield  {author} {\bibinfo {author} {\bibfnamefont {D.}~\bibnamefont
  {Malacara-Hern\'{a}ndez}}\ and\ \bibinfo {author} {\bibfnamefont
  {Z.}~\bibnamefont {Malacara-Hern\'{a}ndez}},\ }\href@noop {}
  {{\selectlanguage {en}\emph {\bibinfo {title} {Handbook of {Optical}
  {Design}, {Second} {Edition}}}}}\ (\bibinfo  {publisher} {CRC Press},\
  \bibinfo {year} {2014})\ \bibinfo {note} {00000}\BibitemShut {NoStop}%
\bibitem [{\citenamefont {Rusz}\ \emph {et~al.}(2014)\citenamefont {Rusz},
  \citenamefont {Idrobo},\ and\ \citenamefont
  {Bhowmick}}]{rusz_achieving_2014}%
  \BibitemOpen
  \bibfield  {author} {\bibinfo {author} {\bibfnamefont {J.}~\bibnamefont
  {Rusz}}, \bibinfo {author} {\bibfnamefont {J.-C.}\ \bibnamefont {Idrobo}}, \
  and\ \bibinfo {author} {\bibfnamefont {S.}~\bibnamefont {Bhowmick}},\ }\href
  {\doibase 10.1103/PhysRevLett.113.145501} {\bibfield  {journal} {\bibinfo
  {journal} {Physical Review Letters}\ }\textbf {\bibinfo {volume} {113}},\
  \bibinfo {pages} {145501} (\bibinfo {year} {2014})},\ \bibinfo {note}
  {00002}\BibitemShut {NoStop}%
\bibitem [{\citenamefont {Pozzi}(1989)}]{pozzi_multislice_1989}%
  \BibitemOpen
  \bibfield  {author} {\bibinfo {author} {\bibfnamefont {G.}~\bibnamefont
  {Pozzi}},\ }\href {\doibase 10.1016/0304-3991(89)90072-7} {\bibfield
  {journal} {\bibinfo  {journal} {Ultramicroscopy}\ }\textbf {\bibinfo {volume}
  {30}},\ \bibinfo {pages} {417} (\bibinfo {year} {1989})},\ \bibinfo {note}
  {00010}\BibitemShut {NoStop}%
\end{thebibliography}%
\end{document}